\documentclass[12pt]{article}
\usepackage[dvips]{graphicx}

\renewcommand{\bar}[1]{\overline{#1}}
\newcommand{\half}{{$\frac{1}{2}$}} 

\begin{document}

\begin{flushright}
SLAC-PUB-9562 \\

\end{flushright}



\def\bege{\begin{equation}}
\def\ende{\end{equation}}

\def\vdp{{d^{4}p}\over{(2\pi)^{4}}}
\def\vdq{{d^{4}q}\over{(2\pi)^{4}}}
\def\half{{1\over 2}}
\def\slash#1{\;\raise.15ex\hbox{/}\kern-.77em #1}
\def\Dslash{\raise.15ex\hbox{/}\kern-.77em D}
\def\lsim{\mathrel{\mathstrut\smash{\ooalign{\raise2.5pt\hbox{$<$}\cr\lower2.5pt\hbox{$\sim$}}}}}

\def\ddp{d\tilde{p}}
\def\a{\alpha}
\def\b{\beta}
\def\D{\Delta}
\def\G{\Gamma}
\def\e{\epsilon}
\def\g{\gamma}
\def\d{\delta}
\def\p{\phi}
\def\vp{\varphi}
\def\r{\rho}
\def\s{\sigma}
\def\l{\lambda}
\def\L{\Lambda}
\def\th{\theta}
\def\om{\omega}
\def\Om{\Omega}

\def\del{\partial}
\def\ha{\frac{1}{2}}
\def\bar#1{\overline{ #1 }}
\def\psibar{\overline{\psi}}
\def\etabar{\overline{\eta}}
\def\sla#1{#1\!\!\!/}
\def\smgroup{U(1)_Y{\otimes}SU(2)_L{\otimes}SU(3)_C}

\def\ra{\rangle}
\def\la{\langle}
\def\lraw{\leftrightarrow}
\def\raw{\rightarrow}
\def\bdlra{\buildrel \leftrightarrow \over}
\def\bdra{\buildrel \rightarrow \over}

\def\eg{{\it e.g.}}
\def\ie{{\it i.e.}}
\def\Dslash{\raise.15ex\hbox{/}\kern-.77em D}
\def\Aslash{\raise.15ex\hbox{/}\kern-.77em A}
\def\Hslash{\raise.15ex\hbox{/}\kern-.77em H}
\def\np#1#2#3{Nucl. Phys. {\bf #1} (#2) #3}
\def\pl#1#2#3{Phys. Lett. {\bf #1} (#2) #3}
\def\prl#1#2#3{Phys. Rev. Lett. {\bf #1} (#2) #3}
\def\pr#1#2#3{Phys. Rev. {\bf #1} (#2) #3}
\def\prd#1#2#3{Phys. Rev. D {\bf #1} (#2) #3}
\def\etal{{\it et al}}

\def\ddp{d\tilde{p}}
\def\ddl{d\tilde{l}}
\def\ddk{d\tilde{k}}
\def\g{\gamma}
\def\mm{m_{\pi}}
\def\ra{\rangle}
\def\la{\langle}
\def\gt{\tilde{\gamma}}
\def\mQbar{m_{\bar{Q}}}
\def\Lraw{\Longrightarrow}
\def\mgx{m_{\tilde{g}}}
\def\ggx{\tilde{g}}
\def\mqx{m_{\tilde{q}}}
\def\qx{\tilde{q}}
\def\mlx{m_{\tilde{l}}}
\def\lx{\tilde{l}}
\def\wx{\tilde{w}}
\def\tx{\tilde{t}}
\def\bx{\tilde{b}}
\def\hpm{{H}^{\pm}}
\def\hpmx{\tilde{H}^{\pm}}

\def\O{{\cal O}}
\def\M{{\cal M}}
\def\tilde{\widetilde}
\def\bar{\overline}
\def\Z{{\bf Z}}
\def\T{{\bf T}}
\def\S{{\bf S}}
\def\R{{\bf R}}
\def\np#1#2#3{Nucl. Phys. {\bf B#1} (#2) #3}
\def\pl#1#2#3{Phys. Lett. {\bf #1B} (#2) #3}
\def\prl#1#2#3{Phys. Rev. Lett.{\bf #1} (#2) #3}
\def\physrev#1#2#3{Phys. Rev. {\bf D#1} (#2) #3}
\def\ap#1#2#3{Ann. Phys. {\bf #1} (#2) #3}
\def\prep#1#2#3{Phys. Rep. {\bf #1} (#2) #3}
\def\rmp#1#2#3{Rev. Mod. Phys. {\bf #1} (#2) #3}
\def\cmp#1#2#3{Comm. Math. Phys. {\bf #1} (#2) #3}
\def\mpl#1#2#3{Mod. Phys. Lett. {\bf #1} (#2) #3}

\def\Lam#1{\Lambda_{#1}}
\def\pf{{\rm Pf ~}}
\font\zfont = cmss10 
\font\litfont = cmr6
\font\fvfont=cmr5
\def\bigone{\hbox{1\kern -.23em {\rm l}}}
\def\ZZ{\hbox{\zfont Z\kern-.4emZ}}
\def\half{{\litfont {1 \over 2}}}
\def\mx#1{m_{\hbox{\fvfont #1}}}
\def\gx#1{g_{\hbox{\fvfont #1}}}
\def\gG{{\cal G}}
\def\lamlam#1{\langle S_{#1}\rangle}
\def\CM{{\cal M}}
\def\CO{{\cal O}}
\def\Re{{\rm Re ~}}
\def\Im{{\rm Im ~}}
\def\lfm#1{\medskip\noindent\item{#1}}



\bigskip\bigskip

\begin{center}
{\Large \bf Physical Renormalization Schemes and Grand Unification}
\end{center}

\vspace{13pt}

\centerline{ \bf Michael Binger and Stanley J. Brodsky}

\vspace{8pt} {\centerline{Stanford Linear Accelerator Center,}}

{\centerline{Stanford University, Stanford, California 94309, USA}}

\centerline{e-mail: binger@slac.stanford.edu and sjbth@slac.stanford.edu}

\bigskip\bigskip

\begin{abstract}

In a physical renormalization scheme, gauge couplings are defined
directly in terms of physical observables. Such effective charges
are analytic functions of physical scales, and thus mass
thresholds are treated with their correct analytic dependence.
In particular, particles will contribute to physical predictions 
even at energies below their threshold. 
This is in contrast to unphysical renormalization schemes such as
$\bar{MS}$ where mass thresholds  are treated as step functions.
In this paper we analyze supersymmetric grand unification in the
context of physical renormalization schemes and find a number of
qualitative differences and improvements in precision over conventional
approaches.  The effective charge formalism presented here provides a template
for calculating all mass threshold effects for any given grand
unified theory. These new threshold corrections may be important in making the 
measured values of the gauge couplings consistent with unification.

\end{abstract}

\newpage



\section{Introduction}

Precision measurements of the gauge couplings and their possible unification
provide some of the few windows to the Planck scale. It is thus important to
have a firm grasp of the theoretical ambiguities involved. This paper attempts
to address some of these ambiguities.

In a physical renormalization scheme, gauge couplings are defined
directly in terms of physical observables. Such effective charges
are analytic functions of physical scales, and thus the thresholds
associated with heavy particles are treated with their correct
analytic dependence.  This is in contrast to unphysical
renormalization schemes such as the $\bar{MS}$ scheme where mass
thresholds  are treated as step functions. In this paper we will
analyze supersymmetric grand unification in the context of
physical renormalization schemes with the goal of systematizing
the effects of light and heavy mass thresholds and improving the
precision of tests of unification compared with conventional
approaches.

In section 2, we motivate physical renormalization schemes with a
simple example and then present the notation and results used
throughout the paper. In section 3, we look more carefully at the
problem of decoupling heavy particles and the errors induced by unphysical
schemes. In section 4, we discuss the canonical self-energy-like
effective charges for the minimal symmetric standard model (MSSM). These
effective couplings run smoothly over spacelike momenta, have non-analytic
behavior only at the expected physical thresholds for timelike momenta, and more
directly measure the strengths of the forces than the charges of
unphysical schemes. The extraction of effective charges from low energy data is
considered. We identify an important modification of the electromagnetic
coupling $\a_{\rm QED}(M_Z)$ due to the proper inclusion of virtual $W^{\pm}$
loops, thus resulting in  a $4\sigma$ change in its numerical value.
Similar modifications are found for the weak mixing angle. As seen
in section 5, these effective charges provide a more natural and
physical framework for examining gauge coupling unification. In
section 5.1, we demonstrate the invalidity of neglecting heavy
threshold corrections in analyzing grand unified models. The more
rigorous treatment of light thresholds in physical schemes gives
rise to new corrections, but these are numerically small for most
sparticle spectra. The treatment of heavy thresholds with various
unification boundary conditions is discussed in section 5.2. In
the simplest scenario, we find that the gauge couplings should
unify at asymptotically large energies and the only heavy
threshold corrections are logarithms of heavy mass ratios,
corrections which can be obtained in unphysical schemes. An
effective unification scale, defined in section 5.3 as the scale
where quantum gravity corrections produce non-negligible
splittings between the gauge couplings, is found to be roughly
$~10^{17}-10^{18}{\rm GeV}$, depending on the specific GUT model
used. Section 5.4 considers more general unification boundary
conditions with finite unification scale. The resulting heavy
threshold corrections are given in Eq.(\ref{epshmore}). This
result combined with the results of section 5.2 may be used to
determine the experimental consistency of any given GUT model.
This is the main result of the paper. An appendix discusses the
details of constructing the effective charges. Throughout our
analysis, we will find several attractive theoretical features of
the supersymmetric regulator, dimensional reduction, which makes
it the preferred regulator for physical effective charge schemes,
even without supersymmetry.

There have been several previous works on threshold effects in
grand unification. In the first such study \cite{ross}, which
appeared just after the discovery of the grand unification, D.A.
Ross uses form factors to define beta functions which are valid
over all energy scales, including near mass thresholds. The
coupling constants run smoothly over all momenta, and nontrivial
threshold corrections are found for grand unification. Despite
this early significant work, most subsequent work on GUTs have
ignored these threshold effects, perhaps due to the complexity of
the Ross approach. 

An exception from the late 1980's is the work of Kennedy and Lynn
\cite{kenlynn}, who defined electroweak effective charges similar 
to the pinch technique charges used in this paper.  

In several papers \cite{kkr} by Kreuzer, Kummer, and Rebhan, the authors 
compared the Vilkovisky-DeWitt effective action (VDEA), the mass-shell
momentum subtraction scheme(MMOM), and Weinberg's effective gauge theory 
(EGT). They wrote down explicit formula for the running charges which
include analytic threshold behavior for all particles. 
In calculating predictions from grand unification,
they assume asymptotic unification at energies much larger than heavy 
particles, so that the only threshold corrections from heavy particles
come from finite constants which are independent of energy scale 
or masses. We find similar results in section 5.2. Furthermore, 
we include the possibility of a finite unification scale in section 5.4,
 which leads to more complicated corrections.

In \cite{farragigrinstein,bmp}, the authors include
the effects of light supersymmetric scalar and fermion thresholds,
although heavy thresholds and gauge bosons virtual effects are not
treated. In \cite{clavellicoultera}, the authors include both
light and heavy threshold corrections, although the treatment of
gauge bosons is not adequate. In Refs. \cite{gilmercader,gil}, the
authors come to several conclusions similar to ours. However,
their definition leads to gauge parameter dependent effective
couplings. 



\section{Physical Renormalization Schemes and Effective Charges}

In order to motivate the re-analysis of supersymmetric unification
given in this paper, we will first discuss some general properties
of renormalization schemes in the presence of massive fields and
determine a criterion for consistent {\it physical renormalization
schemes}. These criteria will not be satisfied by the schemes
conventionally used in unification (and most perturbative
calculations), $\bar{MS}$ and $\bar{DR}$, which have persistent
threshold and matching errors. Heuristically, these errors can be
understood by noting that such schemes implicitly integrate out
all masses heavier than the physical energy scale until they are
crossed, and then they are ``clicked'' on with a step function. Of
course, integrating out heavy fields is only valid for energies
well below their masses. This procedure is problematic since it
does not correctly incorporate the finite probability that the
uncertainty principle gives for a particle to be pair produced
below threshold. Effective charge\cite{Grunberg:1982fw} schemes,
derived from physical observables, naturally avoid such errors and
are formally consistent.

\subsection{A Simple Example}

For the purpose of elucidating the benefits of physical renormalization schemes,
we will give a simple toy example using QED with three fermions,
$e$, $\mu$, and $\tau$.
Consider the amplitude for the process
$e^-\mu^-{\raw}e^-\mu^-$. This can be written as a dressed skeleton
expansion, i.e.
the dressed tree level graph plus the dressed box diagram plus the dressed
double box, etc..
The tree level diagram, dressed to all orders in perturbation theory, is
equal to
the tree level diagram with one modification : the QED coupling
$\a={e^2\over 4\pi}$
is replaced by the Gell-Man--Low--Dyson effective charge
\bege
\a(Q^2) = {\a \over 1 + \Pi_{\g\g}(Q^2)-\Pi_{\g\g}(0)}.
\ende
Hence, from measurements of the cross section, one can measure the
effective charge
at two different scales, $\a(Q_h^2)$ and $\a(Q_l^2)$. Suppose the value of
the electron charge is not known, and we are trying to test the predictions
of QED.
The way to proceed is to use one measurement, say at the low scale $Q_l$,
as an input
to determine $e$. Now the prediction at the high scale $Q_h$ is well
defined, and
represents a test of the theory. More directly, we could just write
$\a(Q_h^2)$ in terms of $\a(Q_l^2)$, leading to the same prediction.
Since the cross section $\sigma_{e^-\mu^-{\raw}e^-\mu^-}(Q^2)$ is
proportional to
$(\a(Q^2))^2$, we are clearly relating one observable to another.
The procedure just outlined is simply an on-shell
renormalization scheme if $Q_l=0$. More generally, we will refer to such a
scheme
as an effective charge scheme, since we are writing a given observable,
here just $\sigma_{e^-\mu^-{\raw}e^-\mu^-}(Q_h^2)$ (or $\a(Q_h^2)$), in terms of
an effective charge, $\a(Q_l^2)$, defined from a measurement of
the cross section at the scale $Q_l$. One could equally well write any
observable in terms
of this effective charge. Note that this approach to renormalization works
for arbitrary
scales, even if the low scale lies below some threshold, say $Q_l<m_{\tau}$,
while $Q_h>m_{\tau}$. Decoupling and the smooth ``turning on'' of the
$\tau$ are manifest.

\begin{figure}[htb]
 \centering \includegraphics[height=3in,angle=90]{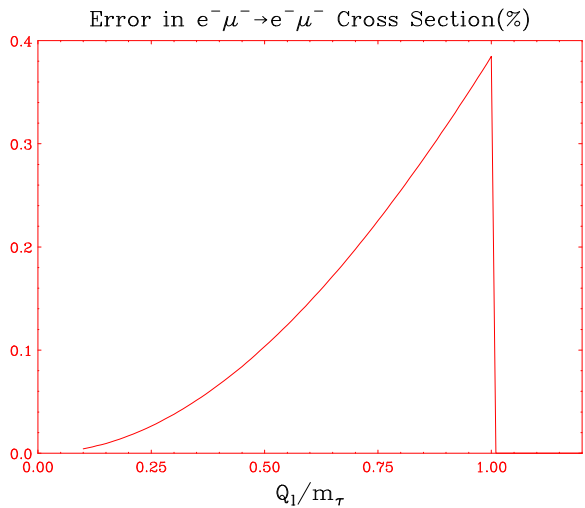}
\caption[*]{The error in the $\bar{MS}$ based prediction for
the scattering cross section,
$
100\%\times\Big( {\hat{\sigma}(Q_h^2)\over \sigma(Q_h^2)} - 1 \Big),
$
plotted against the reference subtraction scale $Q_l$ for the choice
$\a(Q_l)\;\approx\; 0.1$.
}
\label{fig:00}
\end{figure}

Now we will compare with the results obtained by using the
conventional implementation of $\bar{MS}$, which is as follows.
First, the cross section is calculated at $Q_l$ using the rules of
$\bar{MS}$, which allows only the electrons and muons to propagate
in loops, since $Q_l<m_{\tau}$. Comparing the observed cross
section to this result will fix the value of the $\bar{MS}$
coupling for two flavors, $\hat{\a}_2(Q_l)$. To predict the result
of the same experiment at scale $Q_h>m_{\tau}$, we need to evolve
$\hat{\a}_2$ to the tau threshold using the two flavor beta
function, match with a three flavor coupling, $\hat{\a}_3$,
through the relation $\hat{\a}_2(m_{\tau})=\hat{\a}_3(m_{\tau})$,
and then evolve $\hat{\a}_3(m_{\tau})$ to $Q_h$ using the three
flavor beta function. We will now have a prediction for
$\sigma_{e^-\mu^-{\raw}e^-\mu^-}(Q_h^2) \propto (\a(Q_h^2))^2$.
One might expect, from the general principle of RG invariance of
physical predictions that this result should be the same as the
prediction derived using the physical effective charge scheme
above. However, there is a discrepancy arising from the incorrect
treatment of the threshold effects in $\bar{MS}$. A detailed
discussion of this problem will be given in section 3. In any
case, the result can be obtained by straightforwardly applying the
procedure outlined above. One finds that the ratio of the cross
section derived using $\bar{MS}$ with the cross section derived
using effective charges, to first order in perturbation theory, is
given by \bege {\hat{\sigma}(Q_h^2)\over \sigma(Q_h^2)} =
1+2{\a(Q_l)\over 3\pi}\Big( L_{\tau}(Q_l/m_{\tau}) - 5/3 \Big),
\ende
where $L_{\tau}$ is a logarithm-like function (the high energy limit is a
logarithm)
given by
\begin{eqnarray}\label{Ltau}
L_{\tau}(Q/m) &=&\int_0^1 dx 6x(1-x)
\log{\Bigg(1+{Q^2\over m^2}x(1-x) \Bigg)}+5/3 \nonumber\\
          &=&(\b {\rm tanh}^{-1}(\b^{-1}) - 1) ( 3-\b^2 ) + 2,
\end{eqnarray}
where $\b = \sqrt{ 1 + {4m^2\over Q^2} }$.
It satisfies the property $L_{\tau}(0)=5/3$,
so that there is no discrepancy when the low reference scale $Q_l$ is much
lower than
the tau mass threshold. This reflects the important, but often overlooked, fact
that unphysical schemes, such as $\bar{MS}$, are formally consistent only
in desert
regions where particle masses can be neglected. The error is plotted in
Fig.(\ref{fig:00}).
Notice that in this example there is an error only for $Q_l<m_{\tau}$.
However, in the more general case of multiple flavor thresholds, there will be
errors from both high and low scales.
Similar discrepancies will be found in our analysis of grand unification.


\subsection{General Properties of Effective Charges and Physical
Renormalization Schemes}

Effective charges \cite{Grunberg:1982fw}
may be defined for any perturbatively calculable observable
\begin{equation}
{\cal{O}}(Q) \;{\equiv}\; A^{\cal{O}} + a_1^{\cal{O}} \a^{\cal{O}}(Q)
\end{equation}
by absorbing all of the radiative corrections into the effective
charge $\a^{\cal{O}}$. To one-loop order using dimensional
regularization (DREG) or dimensional reduction (DRED), it is
straightforward to show that any unrenormalized effective charge
may be parameterized as\footnote{This follows from considering the
high energy limit and requiring renormalizability. Note also that
our parameterization can be easily extended to effective charges
which have particles with different masses running together in the
loops, and the results are similar. In any case, we will not have
use for such charges in this paper.}
\begin{eqnarray}\label{eqtwo}
\a^{\O}(Q) &=& \a_0-{\a_0^2\over 4\pi}
\sum_p\b_p\Big(  L_p(Q/m_p) - \eta^{\O}_p(Q/m_p)
-C_{UV} + \log{m_p^2\over \mu^2}  \Big) + {\cdots},
\end{eqnarray}
where the sum is over all particles $p$ in the fundamental theory which
contribute
to the running of the effective charge.
In the QED example above, the sum proceeds over $e,\mu,\tau$ and
$L_e=L_{\mu}=L_{\tau}$,
and the function $\eta^{\O}_p(Q/m_p) = 5/3$ is a constant for the simple
observable
${\O} = \sqrt{\sigma_{e^-\mu^-{\raw}e^-\mu^-}}$.
In Eq.(\ref{eqtwo}), $C_{UV} = {1\over \e} -\g_E + \log{4\pi}$ is
the divergence and associated constants, $\mu$ is the regularization scale,
$\a_0$ is the bare gauge coupling, and $\b_p$ is the contribution of each
particle
to the one-loop beta function coefficient. The $L_p(Q/m_p)$ are
logarithmic-like functions which are characteristic of the spin of each
particle, and are given exactly in Eq.(\ref{Lspins}).
They may be approximated to within a few percent\footnote{To be
precise, the approximations reproduce the exact
functions $L_0,L_{1/2},$ and $L_1$ (the subscripts refer to the spin of
the massive field) with maximum error of  $3.5\%$, $0.8\%$, and
$2.2\%$, respectively, over the entire range of $Q$.}
by
\bege\label{Lapprox}
L_p(Q/m)\;{\approx}\;\log{\Big( e^{\eta_p}+{Q^2\over m_p^2} \Big)}
\ende
 and have the limits
\bege\label{Llims}
L_p(Q/m) \;{\buildrel Q{\gg}m \over {\approx}}\; \log{Q^2\over m^2}, \qquad
L_p(Q/m) \;{\buildrel m{\gg}Q \over {\approx}}\; \eta_p,
\ende
where the constants $\eta_p$ have values given in the table below.
We will see that these constants are of central importance in physical
renormalization schemes.
These log-like functions characterize the self-energy-like
effect of each particle, including the finite spread of the wavefunctions
near thresholds due to the uncertainty principle,
and may be calculated in several different ways, as will be discussed in
section 4 and the Appendix. Figure 2 shows the $L_p$ functions for spacelike 
momenta. 

\begin{figure}[htb]
 \centering \includegraphics[height=3.5in,angle=90]{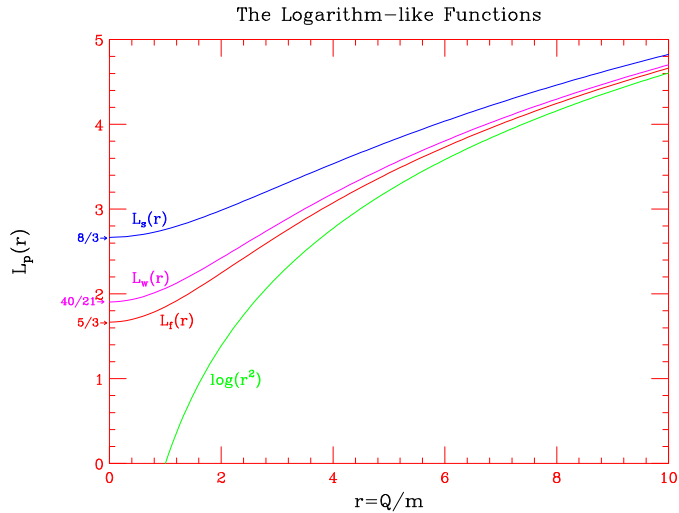}
\caption[*]{The logarithm-like functions for massive particles 
of spin $0, 1/2,$ and $1$ are denoted by $L_s,L_f,$ and $L_W$, respectively. 
}
\label{fig:0.5}
\end{figure}


$$
{\rm Table \;\; I.}
$$
$$
\label{etas}
\def\vspace#1{ \omit & \omit & height #1 & \omit && \omit && \omit &\cr }
\vbox{\offinterlineskip
\hrule
\halign{ \strut
 \vrule#& $\;$  \hfil # \hfil  &\vrule#& $\;$ \hfil #  \hfil &\vrule#&
$\;$ \hfil # \hfil &\vrule#& $\;$ \hfil # \hfil &\vrule#& $\;$ \hfil #
\hfil &\vrule#
\cr
&  && scalars && fermions && massive gauge bosons &\cr
\noalign{\hrule}
\vspace{1pt}
\noalign{\hrule}
& $\eta_p$ && $8/3$ && $5/3$ && $40/21$ &\cr }
\hrule}
$$
The functions $\eta^{\O}_p(Q/m_p)$ are characteristic of each
observable, with a nontrivial functional form indicating
deviations from self-energy like behavior. For a general
observable ${\O}$, the function $\eta^{\O}_p(Q/m_p)$ is
nontrivial. We will show in section 4 that the constants $\eta_p$
correspond to a particularly simple and canonical observable,
called the pinch-technique (PT) self-energy.

The effective coupling renormalized in the most general scheme $R$ is
\begin{eqnarray}\label{robs}
\a^{\O}(Q)&=&
\a_R(Q_0)-{(\a_R(Q_0))^2\over 4\pi}
\sum_p\b_p\Big(L_p(Q/m_p)-L_p(Q_0/m_p)
\nonumber\\&-&
\eta^{\O}_p(Q/m_p)+\eta^R_p(Q_0/m_p)\Big),
\end{eqnarray}
where the functions $\eta^R_p(Q_0/m_p)$ contain all of the
information about the scheme. Here $R$ can be any mathematical
scheme for defining the couplings. In the case of ${\bar{MS}}$, we
have $\eta^{\bar{MS}}_p(Q_0/m_p)=L_p(Q_0/m_p)-\log{(Q_0^2/m_p^2)}$
so that only logarithms of the renormalization scale, $Q_0$, are
subtracted.\footnote{The term in parentheses in Eq.(\ref{robs})
becomes $L_p(Q/m_p) - \log{(Q_0^2/m_p^2)} -
\eta_p^{\cal{O}}(Q/m_p)$. Note that in most calculations the first
term is taken to be a logarithm and mass corrections are
systematically added, in order to approximate the full threshold
dependence of $L_p(Q/m_p)$. However, the $\log{(Q_0^2/m_p^2)}$
term does not have the correct threshold dependence, as we will be
discussing.}

It is straightforward to relate observables to each other:
\begin{eqnarray}\label{ooneotwo}
\a^{\O_1}(Q_1)&=&\a^{\O_2}(Q_2)-{(\a^{\O_2}(Q_2))^2\over 4\pi}\sum_p\b_p
\Big( L_p(Q_1/m_p)-L_p(Q_2/m_p)
\nonumber\\&-&
\eta^{\O_1}_p(Q_1/m_p) + \eta^{\O_2}_p(Q_2/m_p)\Big).
\end{eqnarray}
This satisfies the transitivity property of the physical renormalization group.
As before, the sum over $p$ runs over all particles in the fundamental theory
which contribute to the effective charges.

For consistency, very massive particles must decouple properly and
must not contribute to physical predictions.
Taking the $m_p{\raw}{\infty}$ limit in Eq.(\ref{robs}) and Eq.(\ref{ooneotwo})
yields a fundamental requirement of renormalization schemes and observables:
\bege\label{fundreq}
\eta^R_p(0) = \eta^{\O_1}_p(0) = \eta^{\O_2}_p(0)  \;{\buildrel DRED \over
=}\; \eta_p.
\ende
This consistency requirement
holds for all schemes $R$, observables $\O_1,\O_2$, and for each massive
particle $p$.
These are universal constants for each spin
and, when DRED is used, are equal to the $\eta_p$ given in Table I above,
as can be
verified through explicit calculations. If, for example, DREG was used
instead of DRED,
the last equality of Eq.(\ref{fundreq}) would hold only for fermions and
scalars,
but for spin 1 fields there is an additional constant.\footnote{This is explained below
Eq.(\ref{ptcharge}) and below Eq.(\ref{Lwggtildelim}).}
Renormalization schemes that satisfy (the first two equalities of)
Eq.(\ref{fundreq}) will henceforth
be referred to as physical renormalization schemes, and those that do not
will be called unphysical renormalization schemes, for reasons that will
become clear.

The above discussion implies a unique decoupling limit ($Q/m{\raw}0$)
for observables.
It is interesting that there is also a restriction on the
high energy behavior ($Q/m{\raw}{\infty}$), which holds only for supersymmetric
theories and takes the form of a sum rule. It is given by
\bege\label{sumrule}
{\sum_{p\in S}\b_p(G)\eta_p^{\CO}({\infty}) \over \sum_{p\in S}\b_p(G)} =
K^{\CO},
\ende
where $K^{\CO}$ is a constant that depends
only on the observable, not on the gauge group $G$ or
the supermultiplet $S$. The $\eta_p^{\CO}(Q/m_p)$ are calculated using
DRED, otherwise
the sum rule is true only for differences
$\eta_p^{\CO_1}({\infty})-\eta_p^{\CO_2}({\infty})$ between observables.
Further, the result holds for any number of supersymmetries,
which may be broken or unbroken at low energies.
This can be proven inductively given the result for $N=1$.
It is easy to check using Table I above and the corresponding result
for massless gauge bosons given below Eq.(\ref{ptcharge}) that
$K^{PT} = 2$.
The sum rule just expresses the fact that
there is no resolution within a supermultiplet at high energies, and is
motivated
from conformal invariance and physical renormalization scheme invariance.
Such a sum rule may provide a powerful link between the contributions
of various spin fields to any observable, particularly if a multi-loop or
non-perturbative generalization was found.

\section{Un-Physical Renormalization Schemes and the Problem of Decoupling}

Notice that the physical renormalization scheme requirement
Eq.(\ref{fundreq}) is not met by $\bar{MS}$, $\bar{DR}$,
their massive extensions, or similar schemes. It is well known that
$\bar{MS}$ by itself does not constitute a complete scheme,
rather one must truncate the sum over $p$ to include only
particles with masses less than the scale of the problem. For each
region between thresholds a different scheme is implemented and one must
translate between schemes when crossing thresholds. Hence, $\bar{MS}$ is really
a set of schemes related to each other. We will call such a set an
artificial decoupling scheme(ADS), and now discuss the most general case
at one loop. This will give us an idea of the discrepancies
one may expect in ADS's when compared to the physical renormalization
scheme approach.

Let $S=\{m_1,m_2,...\}$ be the spectrum of massive particles
of the fundamental theory ordered from lightest to
heaviest, let $S_0$ be the set of massless particles,
and let $S^N=S_0{\oplus}\{m_1,m_2,...,m_N\}$ be some subset
up to a given mass scale. For any given renormalization scheme $R$, let the
$N$th phase of $R$, denoted $R^N$, be the scheme used
to renormalize observables at energy scales $Q$ such that
\bege\label{dumbeq}
m_1,...,m_N < Q < m_{N+1}.
\ende
 To use $R^N$, one simply
renormalizes the contributions from particles $p\;{\in}\;S^N$
in the usual way dictated by scheme $R$, and then entirely
neglects the contributions from all $p\;{\in}\;S-S^N$.
This is a formal statement of the usual implementation
of ADS's using step functions.

 For $m_1,...,m_N < Q,Q' < m_{N+1}$, the gauge coupling of the $R^N$ scheme
flows by
\begin{eqnarray}\label{adsflow}
\a_{R^{N}}(Q) &=& \a_{R^{N}}(Q') - ({\a_{R^{N}}(Q'))^2\over 4\pi}
\sum_{p{\in}S^N}\b_p
\Bigg(L_p(Q/m_p)-L_p(Q'/m_p)
\nonumber\\&-&
\eta_p^R(Q/m_p) + \eta_p^R(Q'/m_p) \Bigg),
\end{eqnarray}
and the most general matching condition between schemes $R^{N-1}$ and $R^N$ takes the form
\bege\label{adsmatch}
\a_{R^{N-1}}(\mu_N) = \a_{R^N}(\mu_N)
+{\a_{R^N}^2(\mu_N)\over 4\pi}\b_N A_N(\mu_N/m_N),
\ende
where $A_p(\mu_p/m_p)$ is arbitrary now, but will be specified
below by minimizing errors, and may depend only on the matching
scale $\mu_p$ for each threshold $m_p$, for reasons discussed
below.

For $Q_h>m_{N+n}$ and $m_1,...,m_N < Q_l < m_{N+1}$ (the $h$ and
$l$ stand for heavy and light scales, respectively) one may relate
observables by flowing through $n$ thresholds using the above
formulas to obtain
\begin{eqnarray}\label{adsobsrelation}
\a^{\O_1}(Q_h) &=& \a^{\O_2}(Q_l)-{(\a^{\O_2}(Q_l))^2\over 4\pi}
\Bigg[ \sum_{p{\in}S^{N+n}}\b_p
\Big(  L_p(Q_h/m_p) - L_p(Q_l/m_p) -\eta_p^{\O_1}(Q_h/m_p)
\nonumber\\&+&
\eta_p^{\O_2}(Q_l/m_p) \Big)
+ \sum_{p{\in}S^{N+n}-S^N}\b_p \Big( A_p(\mu_p/m_p) - L_p(\mu_p/m_p)
+\eta_p^{R}(\mu_p/m_p)
\nonumber\\ &+&
L_p(Q_l/m_p)  -\eta_p^{\O_2}(Q_l/m_p) \Big)
\Bigg].
\end{eqnarray}

Now let us compare this to the relation, Eq.(\ref{ooneotwo}),
obtained in the previous section for the tracking of two
observables. For the case of a high scale desert region ($Q_h
{\ll} m_{N+n+1}$), the first sum reduces to Eq.(\ref{ooneotwo}).
However, when $Q_h \lsim  m_{N+n+1}$, there are errors of one loop
order which are proportional to
$L_p(Q_h/m_p)-\eta_p^{\O_1}(Q_h/m_p)$, and occur for each
neglected threshold\footnote{similar errors occur if $Q_l {\lsim}
m_{N+n+1}$} $p{\in}S-S^{N+n}$ such that $Q_h \lsim m_p$. These
errors are naturally remedied in physical renormalization schemes,
leading to heavy threshold corrections which will be of importance
when grand unification is discussed later. There are also
analogous light threshold corrections. The second sum in
Eq.(\ref{adsobsrelation}) contains extra terms which arise from
the artificial decoupling and matching conditions at each
threshold. These terms are also generally of order of the one-loop
corrections and must cancel if the ADS is to be consistent (in the
sense of giving reliable physical predictions in relations between
two observables). In general, these terms do not cancel, since the
$Q_l$-dependent terms cannot be cancelled by the choice of $A_p$,
which depends only on the ratio $\mu_p/ m_p$. Suitably choosing
$A_p$ (see Eq.\ref{adserror} below) leaves a term
$L_p(Q_l/m_p)-\eta_p^{\O_2}(Q_l/m_p)$ for all $p{\in}S^{N+n}-S^N$.
Hence, we see that the high scale and low scale threshold
corrections have exactly the same form; indeed, they have the same
origin, namely the necessity of using an ADS, which arises from
improper decoupling.

In the case of a low scale desert region, $Q_l{\ll}m_p {\forall} p {\in}
S^{N+n}-S^N$,
one finds that $L_p(Q_l/m_p)-\eta_p^{\O_2}(Q_l/m_p) {\raw} 0$ (by
Eqs.(\ref{Llims},\ref{fundreq})),
and the light threshold errors are eliminated through the choice
\bege\label{adserror}
A_p(\mu_p/m_p) =  L_p(\mu_p/m_p) - \eta_p^{R}(\mu_p/m_p).
\ende
 In this case, notice that once the $A_p$ are chosen
suitably, the matching scale $\mu_p$ exactly cancels and there is
no need to fix its value. However, for $\bar{MS}$, this choice is
equivalent to using $A_p=0$ and $\mu_p = m_p$, which is the
matching scale typically used at one loop. One may object that
even in non-desert regions the known anomalous matching threshold
errors could be systematically subtracted off for each physical
process considered (equivalently allowing $A_p=A_p(Q_l)$). This is
untenable, as it is the same as using a different coupling for
each process, thus losing the remnants of universality left by
ADS's (i.e. universality in each desert region).

We have identified two potential problems in artificial decoupling
schemes, which arise solely from the failure of the decoupling
requirement given in Eq.(\ref{fundreq}), regardless of whether or
not the scheme $R$ has analytic threshold
dependence.\footnote{Proper analytic threshold dependence may be
defined by $\eta_p^R(Q/m)$ going to a constant for both small and
large $Q/m$. Consider the analytic extension of $\bar{MS}$ into
the region of mass thresholds, which we call massive $\bar{MS}$,
or $M\bar{MS}$ (similar to that in \cite{mmsbar}). This is defined
by $\eta^{M\bar{MS}}_p(Q_0/m_p) = 0$ so that the full
logarithmic-like functions $L_p$ are subtracted and trivially the
conditions for smooth threshold dependence are satisfied.
Nonetheless, $M\bar{MS}$ has matching errors, which result from
the failure of Eq.(\ref{fundreq}) and the subsequent need to
construct an ADS from $M\bar{MS}$.} The low scale errors come from
the matching conditions and are exhibited in the last two terms of
Eq.(\ref{adsobsrelation}). These are significant only when
$Q_l{\lsim}m_{N+1}$. The high scale errors occur when one is
calculating an observable at an energy $Q_h{\lsim}m_{N+n+1}$ that
is slightly less than masses that should contribute, but are cut
off in an ADS. These two errors will give rise to light and heavy
threshold corrections in unification, as discussed in section IV.

In practice, both types of errors can often be eliminated through a
``threshold shifting'' procedure.
This involves modifying the definition of $R^N$ by replacing Eq.(\ref{dumbeq})
with $m_1,...,m_N < a Q < m_{N+1}$ and making similar subsequent replacements,
and by choosing $a>1$ to be large enough so that the desired thresholds
that are slightly above $Q_l$ or $Q_h$ are `moved' below $a Q_l$, so that no
matching need occur for those thresholds, since they are already implicitly
included in the couplings.
The limit of this procedure as $a{\raw}\infty$ leads to
a formally consistent scheme where no matching or artificial decoupling is used,
but due to the failure of decoupling, it requires inclusion of
contributions from every particle in the (unknown) fundamental theory.
This is the exact situation that caused us to introduce an ADS in the first
place, since we did not want unknown and arbitrarily massive fields contributing
to every physical observable (written in terms of the ADS scheme charge).
It is true that such unknown contributions cancel in relations between
observables,
but the utility of the intermediate ADS scheme is lost since it's coupling is
ill-defined\footnote{We might as well always write observables in terms of other
observables; this is precisely the philosophy of effective charge inspired
physical renormalization schemes.}.
In many calculations in unphysical schemes such as
$\bar{MS}$, the ``threshold shifting''
approach may be used to yield physical predictions which are
arbitrarily accurate by choosing a sufficiently large `a'.
However, the usefulness of this procedure depends on
the details of the mass spectrum. There is no universal
algorithm that applies to any field theory.
The complicated nature of such artificial fixes
to the decoupling problem are reflections of the unphysical nature of
the schemes and couplings. See \cite{msbaralt} for another approach to fixing 
$\bar{MS}$.

Thus, we have shown that the $\bar{MS}$ and $\bar{DR}$ schemes
suffer errors unless one is restricted to observables at energies
$Q$ which lie far between mass thresholds. In addition,
complicated matching conditions must be applied when crossing
thresholds to maintain consistency for such desert scenarios. In
principle, these schemes are only valid for theories where all
particles have zero or infinite mass, or if one knows the full
field content of the underlying physical theory.

\section{The Canonical Physical Effective Charges for the MSSM }

The difficulties associated with unphysical schemes
are circumvented in physical renormalization schemes (PRS) based upon
effective charges. So far, we have given consistency conditions which are
satisfied
PRS's, but explicit examples have not been given. This is the topic we now
take up.

\begin{figure}[htb]
 \centering \includegraphics[height=4in]{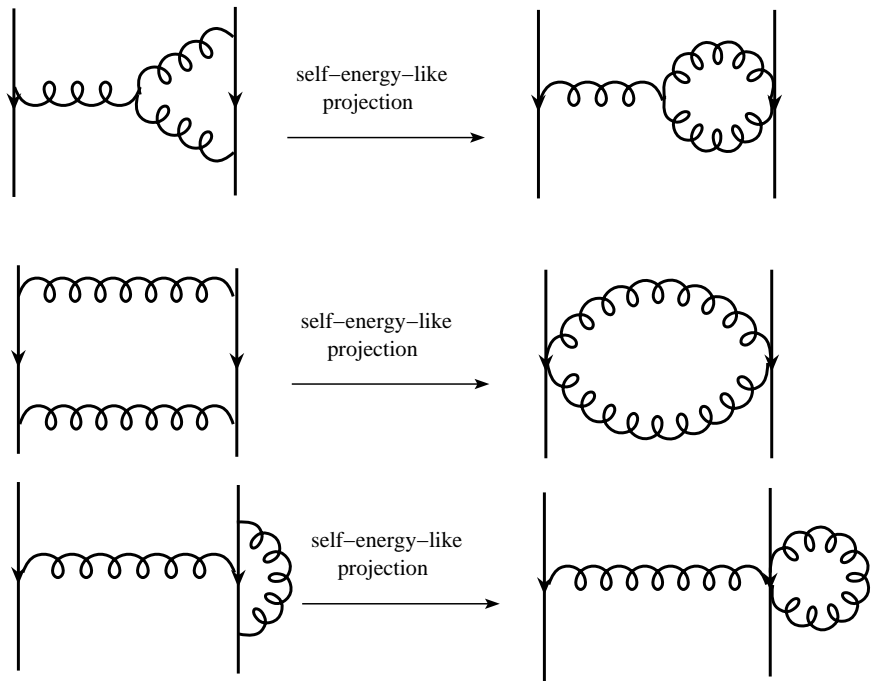} \caption[*]{{\bf
Pinch-technique
for QCD at 1 loop.} The unique gluonic self-energy-like projection
of the vertex and box graphs yield terms which must be added to the conventional
self-energy to get the PT effective charge.} \label{fig:1}
\end{figure}

For any observable ${\cal{O}}$, we define an effective charge scheme
$R^{\cal{O}}$, by
\bege\label{effchschemes}
\eta_p^{R^{\cal{O}}}(Q/m_p)=\eta_p^{\cal{O}}(Q/m_p),
\ende
which, after using Eq.(\ref{robs}),  is equivalent to
\bege\label{effcheq}
\a^{\O}(Q) = \a_{R^{\O}}(Q),
\ende
thus motivating the terminology ``effective charge''. Here
$R^{\O}$ is the physical subset of all possible mathematical
schemes. The canonical example for using an effective charge as a
scheme is furnished in QED by the Gell-Mann--Low--Dyson charge,
which can be measured directly from scattering experiments. The
extension of this concept to non-abelian gauge  theories is
non-trivial\cite{coq}, due to the self interactions of the gauge bosons
which make the usual self-energy gauge dependent. However,
systematically implementing the Ward identities of the theory
allows one to project out the unique self-energy of each {\it
physical} particle, resulting in a self-energy that is gauge
independent, may be resummed to define an effective charge, and
may be related via the optical theorem to appropriate cuts of
differential cross sections. The algorithm for performing the
calculation at the diagrammatic level is called the
pinch-technique (PT)
\cite{cornwall}\cite{degsir}\cite{watson}\cite{prw}. The procedure
is illustrated in Fig.(\ref{fig:1}) for the case of massless gauge
theory, where
 momentum factors from internal gauge boson lines or vertices  combine with
gamma matrices to cancel internal fermion propagators, yielding a gluon
self-energy-like graph. This is then added to the usual self-energy to
yield the full PT self-energy.

The PT procedure is unambiguous at one loop and is merely an
application of the Ward identities of the theory, which becomes
more transparent in a dispersive derivation from physical  cross
sections $\sigma(q\bar{q}{\raw}gg)$ (see \cite{prw} for such a
construction for the electroweak sector). The generalization of
the pinch technique to  higher loops has recently been investigated.
\cite{watson2a}\cite{watson2b}\cite{rafael2}\cite{Binosi:2002ft}
\cite{Binosiqcdall}\cite{Binosiewall}. In the work 
of Binosi and Papavassiliou \cite{Binosi:2002ft}\cite{Binosiqcdall}
\cite{Binosiewall}, the authors prove the consistency of the pinch 
technique to all orders in perturbation theory, making clear how to define 
the QCD and electroweak effective charges at higher orders.

The PT charge, labelled by $\tilde{\a}$, written in terms of the bare coupling
$\a_0$ may be calculated for arbitrary gauge theory, broken or unbroken,
to be\footnote{If particles of different mass propagate together in the loops, this 
formula is trivially modified.}
\bege\label{ptcharge}
{\tilde{\a}}(Q) =
\a_0-{\a_0^2\over 4\pi}
\sum_p\b_p\Big(  L_p(Q/m_p) - \eta_p - C_{UV}+l_m  \Big) + {\cdots},
\ende
where $\eta_p^{PT-DRED}({Q\over m_p}) = \eta_p$ are the constants given in
Table I for massive fields and $\eta_g=64/33$ for massless spin 1
fields\footnote{We will use $'W'$ or $'1'$ subscripts to denote massive
spin 1 fields and
a $'g'$ subscript for massless spin fields. The constants $64/33$ and
$40/21$ are related
straightforwardly. In general, for a massive gauge boson $W$ in the
representation $R$ of group $G$ that is being considered and representations
$R'$ in additional group factors $G'$, we have
\bege
\b_W = {11\over 3}C(R)d(R')-{1\over 6}C(R)d(R')={7\over 2}C(R)d(R')
\ende
and
\bege
\eta_W = {1\over \b_W}\Bigg(
{11\over 3}C(R)d(R') \Big({64\over 33}\Big)
- {1\over 6}C(R)d(R') \Big({8\over 3}\Big)   \Bigg)={40\over 21}.
\ende
}.
The fact that these $\eta_p^{\O}(Q/m)$ functions are constants is what makes
the $PT$ observable
the most simple and  natural choice for defining an effective charge scheme.
More general  physical effective charge schemes
(see Eqs.(\ref{robs},\ref{ooneotwo},\ref{fundreq}))
have more complicated running due to the $\eta_p^{\O}(Q/m_p)$ terms.
The calculation of ${\tilde{\a}}(Q)$ has been performed
using dimensional reduction (DRED), rather than dimensional regularization
(DREG). We will let $\bar{PT}$ stand for the renormalization scheme
associated to
the PT observable regularized using DRED.
The results in DREG for $s=0,1/2$ fields are the same, but for spin one
fields are given by the replacement  $\eta_W\;{\raw}\;\eta_W+2/21$, where the
additional constant is due to the  so-called $\e$ ghosts. Although the
constant terms will cancel  in relations between physical observables, DRED is
the more natural choice, even for non-supersymmetric theories. This is because
decoupling is  manifest (including for massive spin 1 fields) individually
for each observable without relying on cancellation terms arising from relations
to another observable.
This is just a statement of the fact that $\eta_p^{PT-DRED} = L_p(0)
{\forall} p$.
In contrast, with DREG one has $\eta_W^{PT-DREG} \;{\neq}\; L_W(0)$.
Also, the supersymmetric sum rule in Eq.(\ref{sumrule}) becomes manifest in
DRED.

Using the above results, it is straightforward to write down the effective
charges for the standard model through
\begin{eqnarray}\label{smeffch}
\tilde{\a_1}(Q^2) &=& {5\over 3}{\tilde{\a}(Q^2)\over 1-\tilde{s}^2(Q^2)}
\nonumber\\
\tilde{\a_2}(Q^2) &=& {\tilde{\a}(Q^2)\over \tilde{s}^2(Q^2)} \nonumber\\
\tilde{\a_3}(Q^2) &=& \tilde{\a_s}(Q^2),
\end{eqnarray}
where the effective couplings $\tilde{\a}$ and $\tilde{s}^2$ are defined
from PT self-energies $\tilde{\Pi}_{\g\g}$ and $\tilde{\Pi}_{\g Z}$,
respectively \cite{prw}, as is detailed in Appendix A. It is convenient
to write $\tilde{\a_1}$ and $\tilde{\a_2}$ in terms of $\tilde{\a}$ and
$\tilde{s}^2$ since the latter contain the contributions from the mass
eigenstate fields. One could use Eq.(\ref{ptcharge})
directly, although the Higgs sector requires care. 

Several subtleties should be addressed before the numerical values of the
$\bar{PT}$ couplings are given.

An important difference between the physical effective charges and
the unphysical $\bar{MS}$ couplings is a distinction between
timelike and spacelike momenta. In conventional approaches,
thresholds are treated in a step function approximation, and hence
the running is always logarithmic. The analytic continuation from
spacelike to timelike momenta is trivial, yielding $i\pi$
imaginary terms on the timelike side. Thus, the real parts of such
couplings are the same modulo three loop $(i\pi)^2$ corrections.
In contrast, the $\bar{PT}$ couplings on timelike and spacelike
sides have considerable differences at one-loop. To see this we
need the exact expressions for the logarithmic-like functions of a
particle of spin $s$, which can be written as \bege\label{Lspins}
L_s(Q/m) = 2 \Bigg[ \big( \b {\rm tanh}^{-1}(\b^{-1}) - 1 \big)
\Bigg( {4S^2-\b^2\over 4S^2 - 1}  \Bigg) + 1 \Bigg],
\ende
where $S^2 = s(s+1)$ is the total spin squared eigenvalue,
$\b = \sqrt{ 1 + {4m^2\over Q^2} }$, and the momenta is spacelike
$(Q^2 > 0)$. This formula is merely a compact way to write the results for
massive spin $0,1/2,$ and $1$ fields, and has not been explicitly verified
for higher spins. For example, $L_{s=1}$ is calculated
from the sum of the usual gauge boson self interaction loop, the ghost loops,
the appropriate loops of Goldstone bosons that are eaten, and the pinched parts
of the vertex
and box graphs (see appendix A for details). In contrast, $L_{s=1/2}$ is
simply the related to the  usual fermion vacuum polarization graph, and
$L_{s=0}$
comes from the usual scalar contribution to the gauge boson self-energy.
It is interesting that such a simple compact form is obtained,
considering the seemingly different derivations of the three $L_s$ functions.
This may suggest a more efficient formulation of the perturbative dynamics of
quantum fields that treats the various spins in a unified manner
\cite{schwinger}.
Notice that
\bege
\lim_{m{\raw}{\infty}}L_s(Q/m) = {8\over 3}\Bigg[{1-3s(s+1)\over 1-4s(s+1)}\Bigg],
\ende
corresponding to the results of Table I.
The analytic continuation of Eq.(\ref{Lspins}) to timelike momenta below
threshold, $ 0 < q^2 = -Q^2 < 4m^2$, is obtained by replacing
\begin{eqnarray}
\b {\raw} i\bar{\b}, \;\; {\rm where} \;\;
\bar{\b} = \sqrt{{4m^2\over q^2}- 1 }, \;\; {\rm and} \;\;
{\rm tanh}^{-1}(\b^{-1}) {\raw} -i{\rm tan}^{-1}(\bar{\b}^{-1}).
\end{eqnarray}
Above threshold, $q^2 > 4m^2$, one should replace
\begin{eqnarray}
{\rm tanh}^{-1}(\b^{-1}) {\raw} {\rm tanh}^{-1}(\b) + i{\pi\over 2}
\;\;{\rm where}\;\; \b = \sqrt{ 1 - {4m^2\over q^2} }.
\end{eqnarray}
From these results it is clear
that significant differences will arise between the spacelike  and timelike
couplings evaluated at scale $M_Z^2$, mainly due to the $W^{\pm}$ boson
threshold asymmetry.

As has been discussed, another distinction of
effective couplings is that they are automatically sensitive to light
SUSY thresholds near $M_Z$, since the $L_s$ functions
are not zero below threshold (on the spacelike side nor on the timelike side).
The effects of light SUSY thresholds on the values of the couplings
at the Z-pole will depend on the method of extraction from the data.
The key question is whether or not the light sparticles are implicitly
included in the measured values of the couplings at $M_Z$.
For $\tilde{\a}(M_Z)$, which is
extracted by running the precisely known fine structure constant from $Q=0$
to $M_Z$, we should include corrections from virtual effects of sparticles
(with model-dependent mass), in the self-energy term $\Pi_{\g\g}(M_Z)$.
However, these threshold corrections will cancel
in any unification prediction, since then one is essentially running from
$Q=0$ to $Q=M_{GUT}$ and the light SUSYs are either fully decoupled or
fully turned on.
For the strong and weak couplings we use data from the Z-pole,
and thus no unknown sparticle thresholds must be accounted for since
they are already implicitly contained in the measured values. When these
couplings are run to the unification scale the induced light threshold
corrections will not cancel. Of course, linear combinations of the
electromagnetic and weak couplings (Eq.(\ref{smeffch})) are used for
unification, which complicates
the matter further, since different methods of extraction are used for each.
It would be unpleasant to quote a different value of
$\tilde{\a}^{-1}(M_Z)$ for each different SUSY spectra considered. However,
this approach has the advantage that the values of the couplings used
are the values that one would directly measure in an experiment at $M_Z$ if
a given sparticle spectrum were the correct one.
For convenience, we will quote the QED coupling extracted assuming a fully
decoupled SUSY. When calculating detailed unification predictions in given
models,
however, the appropriate terms will be included in the determination of
$\tilde{\a}^{-1}(M_Z)$. It should be emphasized that the above complications
are only numerically significant for light sparticle spectra.

The initial values may be extracted from experimental data
and are given in the following table, where
spacelike and timelike effective couplings are denoted with a $'+'$
and $'-'$, respectively. The $\bar{MS}$ and $\bar{DR}$ couplings
are on the timelike side.

$$
{\rm Table \;\; II.}
$$
$$
\label{effinit}
\def\vspace#1{ \omit & \omit & height #1 & \omit && \omit && \omit &\cr }
\vbox{\offinterlineskip
\hrule
\halign{
\strut
\vrule#& $\;$ \hfil # \hfil &
\vrule#& $\;$ \hfil # \hfil &
\vrule#& $\;$ \hfil # \hfil &
\vrule#& $\;$ \hfil # \hfil &
\vrule#& $\;$ \hfil # \hfil &
\vrule# \cr
& $ $
&& $\bar{MS}$
&& $\bar{DR}$
&& $\bar{PT}_+$
&& $\bar{PT}_-$ & \cr
\noalign{\hrule}
\vspace{1pt}
\noalign{\hrule}
&  ${\a}^{-1}(M_Z)$
&& $127.934(27)$
&& $127.881(27)$
&& $129.076(27)$
&& $128.830(27)$ &\cr
\noalign{\hrule}
&  ${s}^2(M_Z)$
&& $0.23114(20)$
&& $0.23030(20)$
&& $0.23130(20)$
&& $0.22973(20)$ &\cr
\noalign{\hrule}
&  ${\a_3}(M_Z)$
&& $0.118(4)$
&& $0.119(4)$
&& $0.140(5)$
&& $0.140(5)$ &\cr
}
\hrule}
$$

See the appendix for detailed formulas for the effective couplings.

Notice that the value of the $\bar{PT}_+$ electromagnetic inverse coupling,
$\tilde{\a}^{-1}(M_Z) = 129.076(27)$, does not correspond to the usual value
of about $128.968(27)$. This discrepancy arises because $\tilde{\a}^{-1}(M_Z)$
includes the virtual effects of $W^+W^-$ loops, whereas the usual construction
of $\a_{\rm{QED}}(M_Z)$ entirely ignores the virtual effects of the
massive gauge bosons. The proximate cause of this consistent oversight
in the literature is the difficulty in extracting a gauge invariant self-energy-like
contribution to the running couplings for non-abelian theories, a problem
which is resolved through the pinch technique, in particular, and more
generally,
in any effective charge scheme.
Clearly, the present approach yields a coupling which more accurately
reflects the
strength of the electromagnetic force. Similar comments apply to the
weak mixing angle.

It should be emphasized that although we have chosen to discuss a particular
physical renormalization scheme, it will be shown in the next section
that all predictions associated with unification are PRS invariant, as they
should be. However, a definite scheme must be chosen for explicit
calculations, and the $\bar{PT}$ scheme is the simplest choice.
As expected, we will find that PRS invariance does not extend to unphysical
schemes such as $\bar{MS}$ or $\bar{DR}$, because of errors associated with the
incorrect treatment of light and heavy thresholds.

\section{Unification in Physical Renormalization Schemes}

Now we are ready to discuss unification. In section 5.1, we will consider only
the  light spectrum given by the standard model fields and their $N=1$
superpartners. This gives a model-independent starting point for discussing
unification, and makes clear exactly what model dependent heavy threshold
corrections are needed for consistency with the unification hypothesis. New
light threshold corrections, in addition to the usual light
mass corrections, are evident, although they are numerically important
for only a small range of parameter space corresponding to light sparticles.
In section 5.2, {\it asymptotic unification} is introduced, leading to
substantial qualitative changes
in the usual picture of gauge unification. This particular choice of
unification boundary conditions will lead to  corrections from logarithms of
superheavy mass ratios, just as would  be obtained by implementing $\bar{DR}$
with the step function  approximation. This sheds light on the nature of the
approximation of the $\bar{DR}$ approach.
In section 5.3, an effective unification scale is derived that is considerable
higher than the usual unification scale.
In section 5.4, more general non-asymptotic boundary conditions are considered,
and the new non-trivial thresholds corrections are found to be important.

In performing the analysis, the exact analytic one-loop formulas
discussed in section 4 will be used, as well as the leading two-loop
corrections. The analytic mass dependent two-loop corrections are not
known, but these can be estimated to be numerically small and well within the
error bars, and hence can be neglected \cite{mellesbmr}.

We will treat the SUSY spectrum as entirely arbitrary, rather than
assume a particular model or theoretical bias. The advantage of
this approach is that importance of various spectra parameters
becomes transparent, and irrelevant details can be ignored.

\subsection{The (in)validity of Neglecting Heavy Thresholds}
In this subsection only, heavy thresholds will be entirely neglected.

The usual test of unification is to predict $\a_3(M_Z)$ contingent
upon  unification. Compared with the conventional
$\bar{DR}$ framework, we expect to see improvements
due to the correct treatment of light thresholds.
To be precise, the corrections we are discussing are to the difference between
the $\tilde{\a}_3(M_Z)$ prediction obtained from the following two methods:
(a)using the $\bar{PT}_+$ scheme throughout, (b) using $\bar{DR}$ (with the artificial decoupling
and  theta function treatment of light thresholds) to predict
$\hat{\a}_3(M_Z)$, which is then translated to a prediction for
$\tilde{\a}_3(M_Z)$.
Both approaches capture the leading light threshold effects,
which appear as logarithms of light masses.
The additional corrections in the $\bar{PT}$ scheme are from what we will call
{\it analytic light threshold corrections}, since they arise from correctly
and smoothly interpolating between thresholds. These
are largest when there are light supersymmetric  partners near or below $M_Z$.
For most values of the sparticle masses, they fall inside the error bars.
However, such corrections may become more important as the experimental
values of the couplings are determined more precisely. The exact form of the
new corrections will be shown explicitly in section 5.2,
Eqs.(\ref{deltaL},\ref{drbarerror}).

\begin{figure}[htb]
 \centering \includegraphics[height=4in]{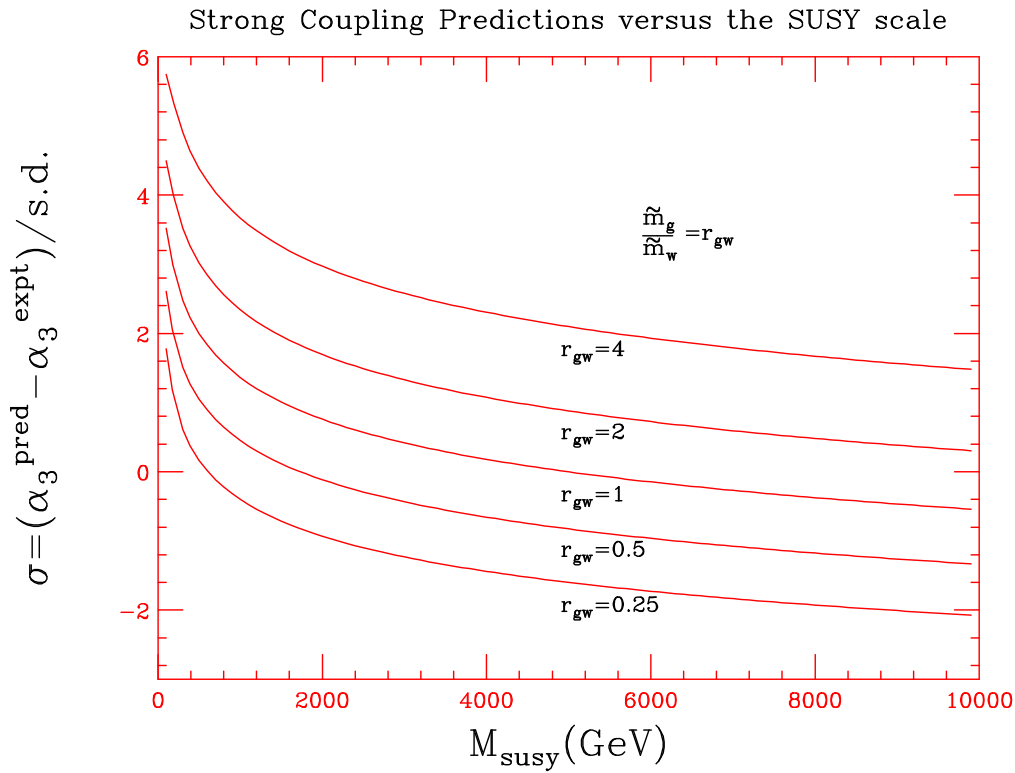} \caption[*]{The error in
the prediction for $\tilde{\a_3}(M_Z)$ is plotted against the
typical SUSY mass scale, with different lines corresponding to values of the
ratio of  the gluino mass to the wino mass.
The relative mass spectrum is roughly the same as most
sparticle spectrum models, including supergravity models, with $M_s$
setting the overall scale.
The experimental standard deviation, s.d., is $0.0055$ for the $PT$ strong
coupling.} \label{fig:2}
\end{figure}

\begin{figure}[htb]
 \centering \includegraphics[height=4in]{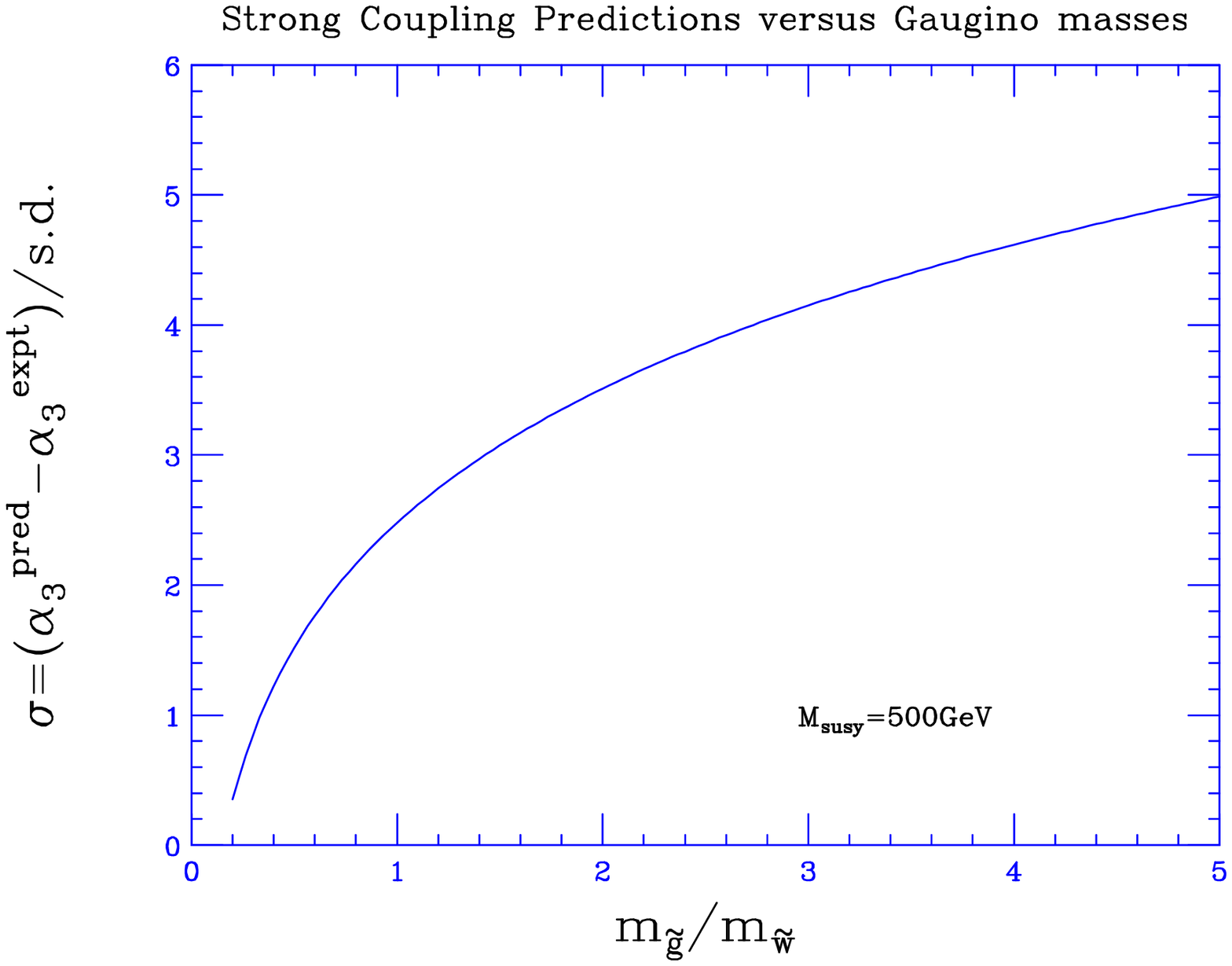} \caption[*]{The error
in the prediction for $\tilde{\a_3}(M_Z)$ is plotted against the
ratio of the gluino mass to the wino mass, which is the sparticle spectrum
parameter  to which $\tilde{\a_3}(M_Z)$ predictions are most sensitive. The
spectrum is  fully specified by the ratio and
$\sqrt{\mgx m_{\wx}}=500{\rm GeV}=M_{\rm susy}$, where $M_{\rm susy}$
is the mass of all other sparticles.} \label{fig:3}
\end{figure}

Now let us compare the $\bar{PT}$ unification predictions with experiment.
The predictions for the $\bar{PT}$ strong coupling, $\tilde{\a_3}$,
(obtained through method (a)) are displayed in Fig.[\ref{fig:2}] against
the SUSY
scale and in Fig.[\ref{fig:3}] against the mass ratio of the gluino and
wino. These are the two SUSY spectrum parameters to which the $\tilde{\a_3}$
prediction is most sensitive.

Only light gluino scenarios with ${\mgx} \;{\lsim}\; m_{\wx}$,
are able to correctly predict the strong coupling for natural SUSY scales
(less than about a TeV).  However, it
is generally expected that the gluino is several times heavier than the
wino for most realistic models of supersymmetry breaking and spectra.
Hence, we reproduce the known result \cite{polonsky} that, at
two loops and neglecting heavy thresholds, gauge coupling unification
fails by several standard deviations. Except for the light gluino escape
route, this points to the need for large heavy threshold corrections
if unification is to be achieved.

\subsection{Heavy Thresholds and Asymptotic Unification}

Henceforth, the complete heavy threshold behavior will be included in the
running of the effective couplings. The form of the subsequent corrections will
depend on the particular unification boundary conditions that are chosen, and
the numerical values of the corrections will depend on the details of the
GUT model. In this section we will choose the simplest boundary conditions,
since it will reproduce known results. Later, more general cases will be
considered.

Generally, there are four parameters which specify the unification
boundary conditions. These are the unification scale, $M_U$, and
the values of the couplings at that scale, $\tilde{\a_i}(M_U)$ for
$i=1,2,3$. For our purposes, we will always assume standard
normalizations and take the couplings to be equal at some scale.
In this case, the only free parameter is $M_U$. The two distinct
cases are for finite $M_U$ and infinite $M_U$. The so-called {\it
asymptotic unification} considered in this section corresponds to
the latter choice, namely $M_U {\raw}{\infty}$ and
$\a^{-1}_1(M_U)=\a^{-1}_2(M_U)=\a^{-1}_3(M_U)$. The asymptotic
unification conditions would be appropriate if the standard model
group $G_{SM}$ is embedded in a simple Lie group $G$ which is
fully restored before gravitational or other string interactions
become relevant, and neglecting any other exotic phenomena. Hence,
this choice is somewhat simple and naive, but it is very
instructive.

 We will find that asymptotic unification reproduces the
same heavy threshold corrections which can be obtained by
unphysical renormalization schemes ($\bar{DR}$) with {\it finite}
unification scale. The reason is that in taking $M_U
{\raw}{\infty}$, one is essentially looking at an observable (the
unification requirement) in a desert region, which, as we have
seen, unphysical schemes are capable of treating without error. At
first sight, it may seem strange that the infinite unification
scale predictions of physical schemes correspond to finite
unification scale predictions of unphysical schemes. However, this
is dictated by the nature of unphysical schemes where masses are
turned on and off with a step function.

\begin{figure}[htb]
 \centering \includegraphics[height=4in]{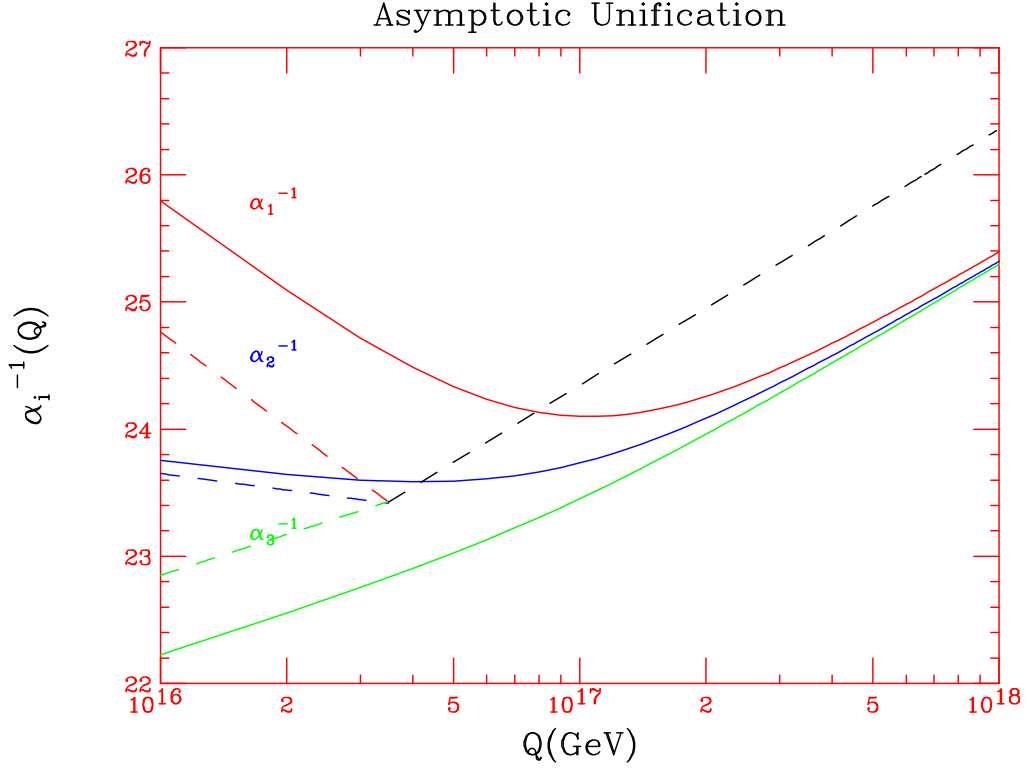} \caption[*]{
{\bf Asymptotic Unification.} The solid lines are the analytic
$\bar{PT}$ effective couplings, while the dashed lines are the
$\bar{DR}$ couplings. For illustrative purposes, $\a_3(M_Z)$ has
been chosen  so that unification occurs at a finite scale for
$\bar{DR}$ and asymptotically for the $\bar{PT}$ couplings. Here
$M_{SUSY}=200{\rm GeV}$ is the mass of all light superpartners
except the wino and gluino which have values ${\half}{\mgx} =
M_{SUSY} = 2 m_{\wx}$. For illustrative purposes, we use $SU(5)$.}
\label{fig:4}
\end{figure}

The paradigmatic improvement over conventional methods is
summarized in Fig.[\ref{fig:4}], where asymptotic unification of the
couplings occurs at very large energy. For demonstrative purposes, the
parameters are chosen so that unification occurs.

Now let us derive the analytic formulae for the unification
predictions. We will discuss the most general case of an $N=1$
supersymmetric $G_{SM}=\smgroup$ embedded in a larger gauge group,
$G$, using any physical RN scheme (all others are inconsistent),
which we label by its associated observable, ${\cal{O}}$.

In general the running of the couplings can be expressed in the
form \bege\label{effrun} \a_{\CO_i}^{-1}(Q) = \a_{\CO_i}^{-1}(Q_0)
+ \tilde{\Pi}^{\CO}_i(Q,Q_0)
                    - \theta_i(Q,Q_0),
\ende
where the two-loop corrections\footnote{see
Appendix A for the details} are contained in $\theta_i(Q,Q_0)$, and we have
defined
\begin{eqnarray}
\tilde{\Pi}^{\CO}_i(Q,Q_0) = {1\over 4\pi}\sum_{p{\in}G}\b_i^{(p)}
\Big(L_p(Q/m_p) - L_p(Q_0/m_p)
- \eta^{\O_i}_p(Q/m_p) + \eta^{\O_i}_p(Q_0/m_p)\Big),
\end{eqnarray}
which contains all of the one loop corrections. Now we separate
the sums over the light and heavy spectra, $L=G_{SM}$ ($G_{SM}$
means the standard model fields plus SUSY partners) and
$H=G-G_{SM}$, take $Q_0=M_Z$, and let $Q=M_U$ be some energy much
larger than the mass of all fields, including the heavy fields;
i.e. $M_U{\gg}m_p\;\;{\forall}p{\in}L+H$. The functions
$\tilde{\Pi}^{\CO}_i$ can then be written as
\begin{eqnarray}
\tilde{\Pi}^{\CO}_i(M_U{\raw}{\infty},M_Z) = \b_G l_U - \Delta_i^L - \delta_i^H-\b_i^H l_X
- S^{\CO}_{L,i}({\infty}) + S^{\CO}_{L,i}\Big( M_Z \Big)-S^{\CO}_{H,i}({\infty})
\end{eqnarray}
where
$\b_G \!=\! \sum_{p{\in}G}\b_p$, $l_U \!=\! {1\over 2{\pi}}\log{M_U \over M_Z}$,
$l_X\!=\!{1\over 2{\pi}}\log{M_X\over M_Z}$,
\bege
S^{\CO}_{L,i}(Q) = \sum_{l\in L} {1\over 4\pi} \b_i^{(l)}
\eta_l^{\CO}(Q/m_l),
\ende
\bege \label{deltaL}
\Delta_i^L = \sum_{l\in L} {1\over 4\pi} \b^{(l)}_i
\Bigg( L_l\Big( { M_Z\over m_l} \Big)
- \log{{M_Z^2\over m_l^2}} \Bigg),
\ende
and
\bege
\delta_i^H = \sum_{h\in H} {1\over 4\pi} \b^{(h)}_i
 \log{{m_h^2\over M_X^2}}.
\ende
The exact 1-loop analytic light threshold corrections are contained in
$\Delta_i^L$,
while the heavy threshold splittings
are contained in $\delta_i^H$, with some arbitrarily chosen heavy mass
$M_X$ which is conveniently taken to be the mass of heavy gauge bosons.

It is useful to verify that predictions
for $l_X$ and $\a_3(M_Z)$ are invariant under the choice of physical
renormalization
scheme. In performing the calculation, one must use the fact that
the $\eta^{\O}_p$ functions do not depend on the gauge group or
representation
of $p$, only the spin. These are necessary (but not sufficient) conditions
for the sum rule
in Eq.(\ref{sumrule}).
This scheme equivalence does not extend to unphysical schemes such as
$\bar{DR}$,
though the errors are quantifiable.

Due to the physical renormalization scheme invariance, we may choose the
simplest scheme, which is the $\bar{PT}$ scheme discussed earlier.
Because the $\eta_p^{\bar{PT}}$ functions are constants equal to $\eta_p =
L_p(0)$,
the expressions for the unification predictions are simple and compact
when written in terms of the $\bar{PT}$ charges $\tilde{\a}_i$.

From $\tilde{\a}_1(M_U)=\tilde{\a}_2(M_U)$,
the heavy gauge boson mass, $M_X$, is given by
\bege\label{mxpred}
{ \log{ \Big( {M_X\over M_Z} \Big) }+1 \over 2\pi} =
{\tilde{\a}^{-1}_2(M_Z)-\tilde{\a}^{-1}_1(M_Z)
+\Delta_{12}\over \b_{12}},
\ende
where $\Delta_{12}=\Delta_{1}-\Delta_{2}$, $\b_{12} = \b_1 - \b_2$, and
$\Delta_{1} = \Delta_1^L + \delta_1^H + \theta_1$.
Notice that $M_X$ can be determined explicitly only for the (unlikely) case
of a degenerate heavy spectrum when $\delta_i^H=0$, otherwise the
expression is transcendental in $M_X$. In the degenerate case, the
gauge boson mass $M_X=M^{\Hslash}_U/e$ is equal to the unification scale
determined by
entirely neglecting heavy thresholds(denoted by ${\Hslash}$), divided by
$e=2.71828...$.
This result relies on use
of the sum rule in Eq.(\ref{sumrule}) which gives rise to the $1/2\pi$ term
on the LHS of Eq.(\ref{mxpred}). The generalization to arbitrary
physical renormalization scheme is $M_X=M^{\Hslash}_U e^{-K^{\O}/2}$,
where $K^{\O}$ is defined in Eq.(\ref{sumrule}).
Neglecting the light and heavy analytic threshold corrections, the
gauge boson mass prediction is the same as the unification scale prediction
in the $\bar{DR}$ scheme. Also, the `unification' scale $M^{\Hslash}_U$
depends on the
particular scheme, which makes sense since different schemes correspond to
different observables. In contrast, $M_X$ is scheme independent.

The strong coupling prediction is
\begin{eqnarray}\label{aspred}
\tilde{\a}_3^{-1}(M_Z) &=&  \tilde{\a}^{-1}_1(M_Z) + \Delta_{31}
+{ \b_{13}\over \b_{12} } \Big( \tilde{\a}^{-1}_2(M_Z)-\tilde{\a}^{-1}_1(M_Z)
+\Delta_{12} \Big),
\end{eqnarray}
which differs from the prediction obtained by
neglecting heavy thresholds by only the terms $\delta_{12}^H$,
$\delta_{13}^H$, which
reflect the heavy splitting.

In order to explicitly compare with the $\bar{DR}$ approach, the
artificial decoupling treatment of thresholds should be employed,
as described in section 2. This involves using a step function
through each light field $l \in L$ with mass greater than $M_Z$,
and through every superheavy field $h \in H$. Then one must impose
the unification condition that the three gauge couplings are equal
at the maximum mass of heavy fields, $M_X = \max{\{ m_h, h \in H
\}}$. At energies above this maximum mass, the three couplings run
identically according to the beta function for the unified group
$G$; hence there is no arbitrariness in the choice of the
unification scale.
Next, the prediction for the $\bar{DR}$ strong coupling  should be
translated to the  $\bar{PT}$ strong coupling. Doing this, one
finds the exact same form of Eq.(\ref{aspred}), except that
$\Delta^L_i$ is replaced by
\begin{eqnarray}\label{drbarerror} \Delta^L_i {\raw} \sum_{m_l<M_Z} {1\over
4\pi} \b^{(l)}_i  \Bigg( L_l\Big( { M_Z\over m_l} \Big)  - \log{{M_Z^2\over
m_l^2}} \Bigg) + \sum_{m_l>M_Z} {1\over 4\pi} \b^{(l)}_i  \Bigg(
\eta_l  - \log{{M_Z^2\over m_l^2}} \Bigg).  \end{eqnarray}
Notice that there are only light threshold corrections beyond the
theta function approximation for particles of mass above $M_Z$,
since those below $M_Z$ are already implicitly accounted for.
This formula is in agreement with Eqs.[\ref{adsobsrelation},\ref{adserror}],
since there is a residual error proportional to $L_p(M_Z/m_p) - \eta_p$
for each crossed threshold. The analogous corrections for the heavy
thresholds do not arise in the asymptotic unification scenario, since we are
essentially comparing  observables at energy scales $M_Z \;{\sim}\; m_l$,
which is of the  same order of magnitude as the light thresholds, and $M_U
{\gg} m_h$, which is much greater than all thresholds when asymptotic
unification conditions are assumed. The latter scale is a ``desert'' scale,
and so the step function method has no errors, giving the same result obtained
above in the $\d^H_i$.\footnote{It should be emphasized that this is only
the case when the $\bar{DR}$ is correctly implemented by choosing the 
unification scale to be equal to the heaviest threshold in the theory.
Different choices are sometimes made in the literature. For example, in 
\cite{lucasraby}, the authors advocate defining the
unification scale to be the geometric mean of the heavy masses.}
For the more general unification conditions considered
in the next subsection there will be additional heavy threshold corrections.

Eq.(\ref{aspred}) is a useful result, as it allows
one to constrain the heavy spectrum, given a light SUSY spectrum.
Up to two loop finite threshold corrections, which we have estimated to be
small,
and assuming that Eq.(\ref{aspred}) will yield the experimental value of the
strong coupling given some appropriate full GUT theory
(i.e. assuming the asymptotic unification hypothesis is true),
we can write
\bege\label{epsh}
\e^H\;{\equiv}\;\tilde{\a}_3^{-1}(M_Z)^{pred}_{\Hslash}-\tilde{\a}_3^{-1}(M_Z)^{expt}
\;{\approx}\;- \delta^H_{31}-{ \b_{13}\over \b_{12} }\delta^H_{12},
\ende
where $\tilde{\a}_3^{-1}(M_Z)^{pred}_{\Hslash}$ is the predicted value of
the strong coupling obtained by neglecting heavy thresholds, as
illustrated in Figs.[\ref{fig:2},\ref{fig:3}]. We should emphasize
the assumptions leading to this result. First, the standard
normalizations of the couplings are assumed, so that
Eq.(\ref{epsh}) does not hold for higher affine levels or
non-standard hypercharge normalizations, as often occur in string
models. Second, we assume that the gluino is somewhat heavier than
the chargino, so that there are serious discrepancies, as in
Fig.[\ref{fig:2}], which must be explained by heavy threshold
corrections. Finally, we are using the paradigm of asymptotic
unification, wherein the full gauge group $G$ in which the SM is
embedded is restored before other Planck scale physics becomes
relevant. With these assumptions, and noting that heavy thresholds
were neglected in Figs.[\ref{fig:2},\ref{fig:3}], we find a
typical value of \bege\label{epsh} \e^H  \;{\buildrel {\rm expt}
\over {\approx}}\; -1 \;{\buildrel {\rm theory} \over {\approx}}\;
-{1\over 4\pi}\sum_{h\in H} B_h \log{m_h^2\over M_X^2},
\ende
where we have defined\footnote{Note that ${ \b_{13}\over \b_{12} } =
{12\over 7}$.}
\bege
B_h  \;{\equiv}\; \b_{31}^{(h)}+{12\over 7}\b_{12}^{(h)}.
\ende
Values of $B^h$ can be compiled for the heavy representations any unified
gauge group, and hence may be used with heavy mass ratios to
exclude or provide evidence for a given GUT theory.

To calculate $B_h$, we first write
$B_h=\bar{\b}_{s_h}\bar{B}_h$, where $\bar{\b}_{s_h} = -1/3,-2/3,11/3$ for spin $0,1/2,1$
fields and the remaining group theory factor is
$\bar{B}_h = {5\over 7}T_1(R) - {12\over 7}T_2(R) + T_3(R)$ for a
representation $R$.
It is necessary to decompose all representations in terms of their
$\smgroup$ content.
Here $T_1(R) = {3\over 5}\sum_{p\in R}Y_p^2$ and
$T_i(R)\d^{ab} = \sum_{p\in R}{\rm tr}_i(t_R^at_R^b), i=2,3$.
For most grand unified theories of interest, all multiplets can be decomposed
in terms of only eight different standard model multiplets
(plus their conjugate representations, which have the same $B_h$,
and a singlet which has $B_h=0$), which are given in
Table III along with the value of $\bar{B_h}$.

$$
{\rm Table \;\; III.}
$$
$$
\label{bh}
\def\vspace#1{ \omit & \omit & height #1 & \omit && \omit && \omit &\cr }
\vbox{\offinterlineskip
\hrule
\halign{
\strut
\vrule#& $\;$ \hfil # \hfil &
\vrule#& $\;$ \hfil # \hfil &
\vrule#& $\;$ \hfil # \hfil &
\vrule#& $\;$ \hfil # \hfil &
\vrule# \cr
& $ $
&& $\bar{B}_h(R_i)$
&&
&& $\bar{B}_h(R_i)$ & \cr
\noalign{\hrule}
\vspace{1pt}
\noalign{\hrule}
&  $R_1 = ({\bf 3},{\bf 2}, 1/6)$
&& $-3/2$
&& $R_5 = ({\bf 1},{\bf 1}, 1)$
&& $3/7$ &\cr
\noalign{\hrule}
&  $R_2 = ({\bf 3},{\bf 1}, -1/3)$
&& $9/14$
&& $R_6 = ({\bf 8},{\bf 1}, 0)$
&& $3$ &\cr
\noalign{\hrule}
&  $R_3 = ({\bf 3},{\bf 1}, 2/3)$
&& $15/14$
&& $R_7 = ({\bf 1},{\bf 3}, 0)$
&& $-24/7$ &\cr
\noalign{\hrule}
&  $R_4 = ({\bf 1},{\bf 2}, 1/2)$
&& $-9/14$
&& $R_8 = ({\bf 3},{\bf 2}, 5/6)$
&& $3/14$ &\cr
}
\hrule}
$$

These same constants will also govern the corrections from analytic heavy
threshold corrections that will be discussed later in section 5.4. Notice
that, by definition,
the constants satisfy the constraint that the sum over all heavy multiplets
vanishes,
\bege
\sum_{h\in H}B_h = 0,
\ende
which equivalently reflects the arbitrariness in the choice of which
heavy mass scale $M_X$ one chooses to be canonical (see Eq.(\ref{epsh})),
${\del \e^H \over \del M_X} = 0$. A similar relation
also holds for any complete representation of the grand unified group.
For example, the ${\bf 24}$ of $SU(5)$ decomposes into a singlet plus
$R_6+R_7+R_8+\bar{R_8}$. From the table, we have
$B_h(R_6) + B_h(R_7) + 2 B_h(R_8)=0$.

As a simple example, let us explore the (unlikely) possibility
wherein the only heavy field with significantly different mass
than the heavy gauge boson mass $M_X$ is the ${\bf 5}$ dimensional
Higgs supermultiplet in which the light Higgs doublets are
embedded. The triplet components of the two Higgs supermultiplets
contributes $-2/5,0,-1$ to $\b_1,\b_2,\b_3$, and hence $B_h({{\bf
3}+\bar{\bf 3}})=-9/7$. Using Eq.(\ref{epsh}), this leads to $M_3
\;{\approx}\; M_X \exp{\Big(-{14\pi \over 9 }\Big)}$, which is of
order $M_X/100$. Such a large splitting is unnatural and difficult
to accommodate in a theory. In general, ``natural'' splittings do
not lead to $\e^H$ values of the correct magnitude in $SU(5)$.
This is not terribly surprising, since minimal SUSY $SU(5)$ is
already known to be strongly disfavored.


In general, the large discrepancies in
Figs.[\ref{fig:2},\ref{fig:3}] imply a large splitting in the
heavy spectrum, which, in turn may imply a multistep unification
scenario, e.g. $SO(10){\raw}G_{224}{\raw}G_{SM}$. The reason is
that for the heavy fields to contribute to $\tilde{\a}_3(M_Z)$, they must
not only have a mass splitting compared to some reference heavy
gauge boson, $X$, but also must have different first beta function
coefficients since only the differences $\b^{(h)}_1-\b^{(h)}_2$
and $\b^{(h)}_1-\b^{(h)}_3$ appear in the corrections.

Before moving to more general unification boundary conditions, we shall
give a simple way to define an effective unification scale in the
asymptotic unification scenario.

\subsection{Effective Unification Scale}

Because the couplings formally unify at infinite energy in the
paradigm of asymptotic unification, there is no apparent
unification scale. However, we suspect that in reality quantum
gravitational fluctuations will affect the couplings as they
approach the Planck energy. Hence, one can define an effective
unification scale to be where the splittings
between the gauge couplings are of the same order as those induced
by gravitational effects. To be precise, define a dimensionless
gravitational coupling which classically runs with energy as \bege
G_N(Q) = {Q^2\over M_{Pl}^2},
\ende
where $M_{Pl} \;{\approx}\; 1.22{\times}10^{19}{GeV}$.
The leading gravitational corrections to the running gauge couplings $\a_i(Q)$
will be proportional to $G_N(Q)\a_i(Q)$. Hence, the effective asymptotic 
unification scale, $M_{eff}$, can be defined as the scale where 
the splittings in the gauge couplings are
of order the gravitational corrections,
$|\a_i(M_{eff}) - \a_j(M_{eff})| \;{\approx}\; b^2 G_N(M_{eff})\a_U$,
or equivalently
\bege
|\a_i^{-1}(M_{eff}) - \a_j^{-1}(M_{eff})| \;{\approx}\; b^2
G_N(M_{eff})\a_U^{-1} \;{\equiv}\; \d_g(M_{eff}),
\ende
where we take $\a_U^{-1} \;{\sim}\; 24$ to be the typical gauge coupling
near unification. The unknown parameter $b^2$ should be of order one.
Estimating $M_{eff}$ using a simple $SU(5)$ model, we find a typical
effective unification scale of $1-5{\times}10^{17}{GeV}$.
This is only intended to a very rough approximation
since a naively simple $SU(5)$ model was used. Nevertheless, more
complicated and realistic GUT models yield a unification scale in
the same ballpark. It is generally true that our effective unification scale
is about an order of magnitude or more greater than what is typically
called the unification
scale (${\sim}2{\times}10^{16}{\rm GeV}$).

It may seem that our definition of an effective unification scale
is rather artificial. However, it may be physically motivated by
the following considerations. If indeed the standard model is
embedded in some unified theory of gravity and gauge forces,
then there may exist a phase at
energies below the Planck scale which consists of a simple Lie
group containing the supersymmetric standard model. In the absence
of any gravitational corrections, the running gauge couplings
certainly unify asymptotically, as this is the only case in which
the higher group symmetry is fully realized up to arbitrarily high
energies. Hence, the running couplings should only deviate from
asymptotic unification by the gravitational corrections
parameterized above. So, by the above reasoning, the effective
unification scale should roughly correspond to the physical
unification scale when the full (quantum gravitational) theory is
considered.

These results may have consequences for the
paradigm of string unification. In particular, one problem of
string unification \cite{dienes} is that the couplings seem to
unify at a scale $(M_G^{\bar{DR}} \;{\approx}\; 2{\times}10^{16}{\rm GeV})$
about twenty times lower than the scale predicted
by four dimensional heterotic string models
$(M_{G}^{string} \;{\approx}\; 5{\times}10^{17}{\rm GeV})$.
In the approach presented here, heavy threshold effects seem
to push the effective asymptotic unification scale to roughly
$M_{G}^{string}$.
Despite the apparent success, this coincidence cannot be taken seriously until
several questions are
addressed in regards to this so-called string gauge coupling problem.
First, the calculation of $M_{G}^{string}$ \cite{kaplunovsky}
was performed in the $\bar{DR}$ scheme, with
the field theory step-function treatment generalized
to strings. An analogous calculation for physical renormalization
schemes is lacking, so it is difficult to compare our results with
string predictions.
Secondly, the asymptotic unification boundary conditions are probably not
valid for many string models, and so the
unification scale will be further changed by more
general boundary conditions, as discussed in the next section. 
See \cite{friedmann} for a recent string calculation of threshold corrections to 
grand unification.


\subsection{More General Boundary Conditions}

The discussion of this section concerns the next-simplest boundary
conditions after asymptotic unification. In particular, we will
impose $\a_1(M_U) = \a_2(M_U) = \a_3(M_U)$ at scale
$M_U {\sim} M_h$, for some $h\in H$. As discussed in
the previous section, one might expect $M_U$ to roughly correspond to the
asymptotic unification scale, which we found to be roughly
$5{\times}10^{17}{\rm GeV}$.
However, we will consider $M_U$ as an input and find the corrections for the
strong coupling and gauge boson mass predictions.

Before proceeding, there is a subtle point that should be
addressed. Notice that in the previous section, we assumed that
unification {\it would have} occurred asymptotically were it not
for gravitational corrections. Hence, starting with a finite
unification scale and then neglecting gravitational corrections,
as we do in this section, does not seem logically consistent with
what was done in the previous section. This observation is
entirely correct, but the point is that indeed we are considering
two orthogonal scenarios, one where a finite unification scale is
obtained from gravity, and another where finite unification scale
is obtained from non-trivial threshold corrections. The latter
case may have its origin in stringy or gravitational physics, but
nevertheless becomes manifest through purely field theoretic
mechanisms.

The corrections from imposing finite unification scale are straightforward
to derive and
can be stated in terms of the
$\Delta_{i} = \Delta_i^L + \delta_i^H + \theta_i$ which we defined earlier.
This gains an additional contribution and can now be written
\bege
\Delta_{i} = \Delta_i^L + \delta_i^H + \theta_i - \Delta_i^H,
\ende
where
\bege
\Delta_i^H = \sum_{h\in H} {1\over 4\pi} \b^{(h)}_i
\Bigg( L_h\Big( { M_U\over m_h} \Big) - \log{{M_U^2\over m_h^2}} \Bigg),
\ende
which is of exactly the same form expected from
Eq.(\ref{adsobsrelation})\footnote{This is not obvious; one must
work through the derivation to see that indeed the expected
$L_p-\eta_p$ correction terms do arise.}. Evidently, these are
finite heavy threshold corrections in addition to the corrections
from the heavy threshold splittings. Hence, the $\e^H$ defined
earlier will get an additional contribution from the
$\Delta^H_i$'s and is now
\begin{eqnarray}\label{epshmore}
\e^H  &\;{\buildrel {\rm theory} \over {\approx}}\;&
{1\over 4\pi}\sum_{h\in H} B_h \Bigg[ \log{m_h^2\over M_X^2} -
\Bigg( L_h\Big( { M_U\over m_h} \Big) - \log{{M_U^2\over m_h^2}} \Bigg)\Bigg]
\nonumber\\
&=& -{1\over 4\pi}\sum_{h\in H} B_h L_h\Big( { M_U\over m_h} \Big).
\end{eqnarray}
Experimentally, $\e^H {\approx} -1$, as seen in
Figs.[\ref{fig:2},\ref{fig:3}] for typical
gluino to wino mass ratios; this value can be easily adjusted for nonstandard
sparticle spectra. This is our final formula which may be used to assess
the experimental validity of gauge coupling unification in any specific GUT
model where the gauge group, superheavy mass ratios, and light SUSY masses
are given.

Let us now consider the numerical size of these new threshold corrections.
From Eq.(\ref{Lapprox}) one finds that
\bege
L_h\Big( { M_U\over m_h} \Big) - \log{{M_U^2\over m_h^2}} \;{\approx}\;
\log{\Big( 1+{m_h^2 \over M_U^2}e^{\eta_h} \Big)},
\ende
which can be larger than heavy splitting corrections $\log{m_h^2\over M_X^2}$
for values of $M_U$ that are not too large. Hence, such corrections cannot
be neglected.

The value of $M_U$ is not fixed a priori, and corresponds to the
physically meaningful energy where the couplings become equal due
to the new nontrivial heavy threshold corrections. This
complicates the analysis of unification by introducing another
parameter beyond those that are usually needed. However, this is
to be expected, since a new physical phenomena (corrections
arising from the virtuality of very massive particles) has been
included.

\section{Conclusions}

We have developed a new way of looking at detailed predictions of
gauge coupling unification which is more physically motivated than conventional
approaches. In addition to a dramatic paradigmatic improvement, novel heavy
and light
threshold corrections are obtained, and the resulting corrections to unification
predictions are presented for a general GUT model.
A natural extension of this work is a thorough analysis
and classification of various unified theories. By calculating the
$B_h$ constants and the heavy spectrum, one may exclude or verify the
gauge unification of a given model.



\begin{center}{\large Acknowledgements} \end{center}

We wish to thank Ratin Akhoury and Eduardo de Rafael for reading a draft of this 
paper and providing useful suggestions.

\begin{center}
{\Large {\bf Appendix A : The Pinch Technique Self-Energy-Like 
Effective Charges} }
\end{center}
\vspace{0.4 cm}

Here we will give explicit formulae for the pinch technique effective
couplings regularized using dimensional reduction (DRED), which will be 
denoted with a tilde.
These effective charges will be similar to those constructed
in \cite{cornwall}\cite{watson} for QCD, and in \cite{degsir}\cite{prw}
for the electroweak sector.
However,  we will  extend these  to the  minimal  supersymmetric case,
which involves  explicitly including another Higgs  doublet,
and regulating the loop integrals with
dimensional reduction (DRED), as opposed to dimensional regularization
(DREG), which is used in most non-supersymmetric settings.
It is well known that DRED preserves both supersymmetry and gauge symmetry.
Also,  the effective charges presented in \cite{prw} were
in  the  on-shell  subtraction  scheme ($Q_0 = 0$),
whereas  here  we  will need the result for arbitrary renormalization scale.
In the appropriate limits our results
reduce  to  those  given  in \cite{watson}  and  \cite{prw}.

The  charges  are constructed using the  pinch-technique (PT), which allows
one to extract the universal self-energy function in non-abelian gauge
theories, thus leading to gauge invariant effective couplings which
\begin{itemize}
\item
 contain explicit and complete mass-threshold behavior and
\item
 reproduce the conventional  massless beta  function in  the limit
where masses can be neglected.
\end{itemize}
At one-loop, the spin $1/2$ and spin $0$ PT self-energies are trivially
just the usual transverse vacuum polarization graphs.
Only the graph with a gauge boson loop needs to have the self-energy-like part
projected, as described briefly in section III, and in more detail in
the references \cite{watson}\cite{prw}. In calculating the following,
we used both the direct diagrammatic pinch technique algorithm\cite{watson}
and the dispersive derivation from physical cross sections\cite{prw}.

The PT effective charges naturally measure the self-energy-like
propagation of a gauge boson and
hence can be interpreted as measuring the real force between two
fermions of arbitrary mass, analogous to the QED
effective charge.
The PT charge includes finite mass recoil effects that
are missed in the heavy quark effective charge (the V-scheme).
In fact, one may obtain the heavy quark potential in the appropriate
kinematical limit of the pinch technique effective charge \cite{watson}.
The difference between the two are due to finite
mass test-charge effects that are not present in the (V) charge but are in
the (PT) charge.  Different extensions of the PT effective charge beyond
one-loop have been put forth  \cite{watson2a}\cite{watson2b}\cite{rafael2},
although it seems that the approach of \cite{rafael2} most closely matches the
philosophy used here. A multi-loop generalization of this algorithm
remains to be constructed.

\vspace{0.4 cm}

{\Large{\it QCD Effective Charges}}

\vspace{0.4 cm}

The $\bar{PT}$ self-energy function for supersymmetric QCD, $\tilde{\Pi}_{3}$,
can be used  to  define  the  effective  coupling  for supersymmetric QCD by
\bege\label{effathree}
\tilde{\a}_{3}(Q) = {\tilde{\a}_{3}(Q_0) \over 1
+ \tilde{\Pi}_{3}(Q,Q_0)}.
\ende

The function $\tilde{\Pi}_{3}$ can be written down straightforwardly using
Eqs.[\ref{eqtwo},\ref{ooneotwo}], and the unsubtracted result is given by
\begin{eqnarray}\label{pithree}
\tilde{\Pi}_3(Q) &=& {\tilde{\a}_{3}(Q_0) \over 4\pi} \Bigg[
  {11\over 3} N_c \Big( \log{Q^2\over \mu^2} - C_{UV} -64/33 \Big)\nonumber\\
&-& {2\over 3} N_c \Big( L_{\half}(Q/m_{\tilde{g}}) +
\log{(m_{\tilde{g}}^2/\mu^2)} - C_{UV} - 5/3 \Big) \nonumber\\
&-&\sum_q  {2\over 3} \Big(  L_{\half}(Q/m_q) + \log{(m_q^2/\mu^2)}-C_{UV}-5/3
\Big) \nonumber\\
&-& \sum_{\qx} {1\over 3} \Big(  L_0(Q/m_{\qx})+
\log{(m_{\qx}^2/\mu^2)}-C_{UV}-8/3 \Big)\Bigg].
\end{eqnarray}

The four terms correspond respectively to the gluons, gluinos
($\tilde{g}$), Dirac quarks ($q$), and to complex squark doublets($\qx$). 
For  the  scalars
we will take the left and right components to be  degenerate in
mass since such complications do not change the unification
predictions to any numerical significance. In any case, one may
trivially treat the two separately.

To relate the resulting effective charge to other schemes or observables
one needs to use Eq.(\ref{ooneotwo}).

 Eq.(\ref{pithree}) can be written in a more useful once subtracted form by
relating the effective charges at different scales, leading to an expression
governing the running of the charge given by
\begin{eqnarray}\label{pithreeren}
\tilde{\Pi}_3(Q,Q_0) &\equiv& \tilde{\Pi}_3(Q) - \tilde{\Pi}_3(Q_0) \nonumber\\
&=& {\tilde{\a}_{3}(Q_0) \over 4\pi} \Bigg[
  {11\over 3} N_c \Big( \log{Q^2\over Q_0^2} \Big)
- {2\over 3} N_c \Big( L_{\half}(Q/m_{\tilde{g}})
- L_{\half}(Q_0/m_{\tilde{g}})\Big) \nonumber\\  &-&\sum_q  {2\over 3}
\Big(  L_{\half}(Q/m_q) - L_{\half}(Q_0/m_q) \Big)\nonumber\\
&-& \sum_{\qx} {1\over 3} \Big(  L_0(Q/m_{\qx}) - L_0(Q_0/m_{\qx}) \Big)\Bigg].
\end{eqnarray}

 Though the gluon contribution in Eq.(\ref{pithree}) looks simple, it is
actually the  most difficult piece to compute. As discussed in
\cite{watson}, the pinch technique self-energy that contributes to
the effective charge is gauge and scale independent, and indeed
reproduces the pure gauge term of the $\b$ function coefficient.
This of course is not the case for  the full pure gluon vacuum
polarization, which is gauge and scale dependent and does not
reproduce the correct  $\b$ function. The non-trivial and
important part of the result is the constant $64/33$. This
constant may be obtained using the pinch technique and DRED. To
translate to DREG one just subtracts $1/11$ (from the so-called
epsilon ghosts) to get the constant $67/33$. For comparison, the
heavy quark potential effective charge, $\a_V$,  replaces this
constant with $28/33$ and $31/33$ when using DRED and DREG,
respectively. Consequently, the V-scheme doesn't satisfy the
decoupling criterion of Eq.(\ref{fundreq}). This is just a
reflection of the fact that infinitely heavy external quarks are
used in the V-scheme calculation, thus rendering meaningless the
limit where internal virtual gauge bosons acquire very large mass.

Notice that in the appropriate limit the above reduces to the standard
RG $\b$ function coefficient,
\bege\label{limpithree}
\lim_{m_i{\raw}0} \tilde{\Pi}_{3}
= {\a_s \over 4\pi}(9-n_f)\log{Q^2\over Q_0}.
\ende

\vspace{0.4 cm}

{\Large {\it The Electroweak Sector}}

\vspace{0.4 cm}

The effective QED charge and the effective weak-mixing angle are obtained by 
diagonalizing the electroweak neutral currents and are given by \cite{prw}
\bege\label{effa}
{\tilde{\a}}(Q) = {{\tilde{\a}}(Q_0)\over 1 + {\tilde{\Pi}}_{\g\g}(Q,Q_0)},
\ende
and
\bege\label{effsin}
\tilde{s}_w^2(Q) = \tilde{s}_w^2(Q_0)
\Bigg( 1 + { \tilde{c}_w(Q_0) \over \tilde{s}_w(Q_0)}
{ {\tilde{\Pi}}_{\g Z}(Q,Q_0)\over 1 + {\tilde{\Pi}}_{\g\g}(Q,Q_0) } \Bigg).
\ende

For the matter sector, we will write only the subtracted $\bar{PT}$
self-energies, as it is now clear how to translate between schemes using the
$\eta_p$ constants as described before.
The quarks ($q$) and leptons ($l$),
along with their scalar superpartners ($\qx,\lx$), yield
\begin{eqnarray}\label{pigg}
\tilde{\Pi}_{\g\g} (matter) &=& {{\tilde{\a}}(Q_0)\over 4\pi} \Bigg[
-\sum_{q} {4\over 3} N_c e_q^2  \Big( L_{\half}(Q/m_q) - L_{\half}(Q_0/m_q)
\Big)
\nonumber\\
&-&\sum_{\qx} {2\over 3} N_c e_{\qx}^2 \Big( L_0(Q/m_{\qx}) -
L_0(Q_0/m_{\qx}) \Big) \nonumber\\
&-&\sum_{l} {4\over 3} \Big( L_{\half}(Q/m_l) - L_{\half}(Q_0/m_l) \Big)
\nonumber\\
&-&\sum_{\lx} {2\over 3} \Big( L_0(Q/m_{\lx}) - L_0(Q_0/m_{\lx}) \Big)
\Bigg].  \end{eqnarray}
The electric charge of a particle $p$ is denoted $e_p$.
The analogous contribution of individual Dirac mass eigenstate 
matter fields to the $\g Z$ self  energy are given  by the relation
\bege\label{pigzpigg}
\tilde{\Pi}^{(p)}_{\g Z} = \Big( {1\over 4|e_p|} - \tilde{s}_w^2 \Big)
{1\over \tilde{s}_w \tilde{c}_w}
\tilde{\Pi}^{(p)}_{\g\g},
\ende
where $p$ denotes any of the fermions or scalars above.

The  contribution   of  the  charged  vector  bosons   to  the
self-energies is more complicated than the matter multiplets. Similar to
the  QCD case,  the  non-abelian  nature of  the  theory implies  that
$W^+W^-$  loops  (along  with   possible  gauge  dependent  ghost  and
Goldstone boson loops),  do not yield a gauge  invariant
result, and do not give the appropriate contribution to the electroweak beta
functions.  The  proper treatment involves  calculating the self-energy
like part of the  one-loop $e^+e^- {\raw} e^+e^-$ amplitude
(or using any other fermions due to universality), including
vertex and  box corrections involving neutrinos. These contribute pinched
parts which make the self-energy-like part gauge invariant and
transverse. This
calculation was  first performed in \cite{degsir}, and then with dispersion
relations in \cite{prw}, for $n_H=1$  Higgs doublets and renormalized in the
on-shell scheme at $Q_0=0$. Here we need to extend these results to
arbitrary $n_H$ and $Q_0$, and would like to have the finite constants in
the unrenormalized expression, including constant terms arising from
using DRED instead of DREG.  The most efficient way to  do this is
to use  the Feynman gauge  $\xi=1$, where $W^{\pm}$  bosons, $-G^{\pm}$
Goldstone    bosons,    and    $\eta^{\pm}$ ghosts  all propagate with
$-ig_{\mu\nu}/(p^2-M_W^2)$. Hence, the only factors of transverse
momenta arise from the three boson vertex, and so the box graph and
several of the vertex graphs may be neglected, as they do not
have pinched parts.
Here, the dependence on the Higgs doublets
comes  only from  the  unphysical charged  Goldstone  scalars for  the
$n_H=1$ case, and also on charged Higgs for $n_H>1$.  The result for the
$SU(2)_L{\otimes}U(1)_Y$ electroweak theory is
\begin{eqnarray}\label{piggwha}
\tilde{\Pi}_{\g\g}^{(W,H)} &=& {{\tilde{\a}}(Q_0)\over 4\pi}\Bigg[
2{11\over 3}\Big(\bar{L_1}(Q/M_W)
+\log{(M_W^2/\mu^2)}-C_{UV} -64/33 \Big)\nonumber\\
&+&\sum_{a=1}^{n_H}\Big(-{1\over3}\Big)
\Big(L_0(Q/M_a)+\log{(M_a^2/\mu^2)}-C_{UV} -8/3 \Big)
\Bigg],
\end{eqnarray}
where the constant  $64/33$ is the same as appeared  for the gluon self
energy. The sum in the second line will be over mass eigenstate charged 
Higgs scalars; there will be one of these for each Higgs doublet in the theory. 
The first scalar ($a=1$ in the sum) is an unphysical Goldstone boson
that is eaten by the $W^{\pm}$, and hence one identifies its mass to be
$M_1=M_W$ (in the Feynman gauge). The second charged scalar (a=2) is
conventionally denoted by $H^{\pm}$ in the MSSM, with mass $M_2=M_{H^{\pm}}$.
Additional Higgs doublets beyond the MSSM are not considered here so we can
take $n_H=2$. The function
\bege\label{Lwggtilde}
\bar{L_1}(Q/m)   =  {2\over   11}\Big(   \b{\rm  tanh^{-1}}
{\b^{-1}}(12-\b^2)+\b^2-1 \Big),
\ende
with $\b = \sqrt{ 1 + {4m^2\over Q^2} }$, 
comes from the $W^+W^-$ and ghost loops,
the $W^+G^-+W^-G^+$ loops, and the pinched self-energy-like part
of the $\g WW$ vertex where the internal neutrino line is pinched.
The $L_0$  comes from the charged Goldstones and Higgs loops.
As might be anticipated from the fermions and scalars, where,
for example, $\lim_{m{\raw}{\infty}}L_{1 \over 2}(Q/m)=5/3$ is the same
constant as appears in the self-energy, we also
have the nice property that
\bege\label{Lwggtildelim}
\lim_{M_W{\raw}{\infty}}\bar{L_1}(Q/M_W)=64/33.
\ende
Notice that in DREG this does not cancel the constant, which in that case is
$67/33$. With DRED regularization, all massive particles decouple, modulo
divergent pieces, from the unsubtracted self-energy-like expression.

Letting the W-bosons eat the Goldstones by performing simple algebra in
Eq.(\ref{piggwha}), one finds the result written in terms of physical
degrees of freedom,
 \begin{eqnarray}\label{piggwhc}
\tilde{\Pi}_{\g\g}^{(W,H)} &=& {{\tilde{\a}}(Q_0)\over 4\pi}\Bigg[
7\Big(L_1(Q/M_W)+\log{(M_W^2/\mu^2)}-C_{UV} -40/21\Big)\nonumber\\
&+&\sum_{a=2}^{n_H}\Big(-{1\over3}\Big)
\Big(L_0(Q/M_a)+\log{(M_a^2/\mu^2)}-C_{UV} -8/3 \Big)
\Bigg],
\end{eqnarray}
The contribution of the physical massive gauge boson is characterized
by $7L_1=(22/3)\bar{L_1}+(-1/3)L_s$, explicitly given by
\bege\label{Lwgg}
L_1(Q/m) = 2\b{\rm tanh^{-1}} {\b^{-1}} \big( 1-(\b^2-1)/7 \big)+(2/7)(\b^2-1).
\ende
As expected,
\bege\label{Lwggqlim}
\lim_{m{\raw}0}L_W(Q/m) = \log{Q^2\over m^2}
\;\;\;\;\;\;\; \lim_{m{\raw}{\infty}}L_W(Q/m) = 40/21.
\ende
Notice that Eq.(\ref{Lwgg}) precisely corresponds to Eq.(\ref{Lspins}) for
$s=1$.

  The separation of pure gauge
effects and those arising in the broken phase of the theory is useful,
and allows us  to immediately write down the  analogous result for $\g
Z$ without further calculation :
\begin{eqnarray}\label{pigzwha}
\tilde{\Pi}_{\g Z}^{(W,H)} &=& {{\tilde{\a}}(Q_0)\over 4\pi c_ws_w}\Bigg[
2{11\over 3}c_w^2\Big(\bar{L_1}(Q/M_W)+\log{(M_W^2/\mu^2)}-C_{UV} -64/33 \Big)
\nonumber\\
&+&\sum_{a=1}^{n_H}\Big(-{1\over 3}(c_w^2-\half)\Big)\Big( L_0(Q/M_a)+
\log{(M_a^2/\mu^2)}-C_{UV} -8/3 \Big)
\Bigg].
\end{eqnarray}

Finally, the wino  and  charged  Higgsino, whose mixing is neglected,
contribute
\bege\label{piggwxhx}
\tilde{\Pi}_{\g\g}^{\tilde{H},\tilde{W}} =
{{\tilde{\a}}(Q_0) \over 4\pi}\Bigg[ -{4\over 3}
 \Big( L_{\half}(Q/m_{\tilde{H}}) - L_{\half}(Q_0/m_{\tilde{H}}) \Big) -
{4\over3} \Big( L_{\half}(Q/m_{\tilde{W}}) - L_{\half}(Q_0/m_{\tilde{W}})
\Big) \Bigg]
\ende
and
\bege\label{pigzwxhx}
\tilde{\Pi}_{\g Z}^{\tilde{H},\tilde{W}} =
{{\tilde{\a}}(Q_0)\over 4\pi c_w s_w}\Bigg[ -{4\over 3}c_w^2 \Big( L_{\half}(Q/m_{\tilde{H}})
- L_{\half}(Q_0/m_{\tilde{H}}) \Big) -{4\over 3}(c_w^2-1/2)
\Big( L_{\half}(Q/m_{\tilde{W}}) - L_{\half}(Q_0/m_{\tilde{W}}) \Big) \Bigg].
\ende

The $SU(2)_L{\otimes}U(1)_Y$ effective couplings constructed from the above results are   
\bege\label{effi}
{\tilde{\a_i}}(Q) = {{\tilde{\a_i}}(Q_0)\over 1 + {\tilde{\Pi}}_{i}(Q,Q_0)},
\ende
for $i=1,2$ and ${\tilde{\a_i}}(Q)$ given in Eq.(\ref{smeffch}). The PT self-energies 
are related by 
\begin{eqnarray}\label{selfenergies}
{\tilde{\Pi}}_{1}&=& {\tilde{\Pi}}_{\g\g}-{s_w\over c_w}{\tilde{\Pi}}_{\g Z}  \nonumber\\
{\tilde{\Pi}}_{2}&=& {\tilde{\Pi}}_{\g\g}+{c_w\over s_w}{\tilde{\Pi}}_{\g Z}.  
\end{eqnarray}

Notice that  $\tilde{\Pi}_{\g\g} = \sum_p
\tilde{\Pi}_{\g\g}^{(p)}$ and $\tilde{\Pi}_{\g Z} = \sum_p
\tilde{\Pi}_{\g  Z}^{(p)}$ (as well as $ \tilde{\Pi}_{i} $ ) 
have the correct  beta function
coefficients, which are summarized below, and   smoothly
interpolate   between  all   mass   thresholds. The  full
mass-dependent beta  functions may  be  obtained by
differentiating the above expressions, but we will just give the
massless limits, in order to make clear our conventions.

The one-loop beta function coefficients are defined by the relations
\bege
{d\a\over d\log{Q^2}} = -\a{d\tilde{\Pi}_{\g\g}\over d\log{Q^2}}
= -{\a^2\over 4\pi}\b_{\g\g}
\ende
\bege
{d  s_w^2  \over  d\log{Q^2}}  = s_w c_w  {d\tilde{\Pi}_{\g  Z}\over
   d\log{Q^2}} = {\a \over 4\pi}\b_{\g Z}
\ende
\bege
\b_1 = {3\over 5} (c_w^2\b_{\g\g}-\b_{\g Z})
\ende
\bege
\b_2 = s_w^2\b_{\g\g}+\b_{\g Z}.
\ende
One finds
\begin{eqnarray}
\b_{\g\g} &=& -{16\over 3} N_g + 6 - n_H \nonumber\\
\b_{\g Z} &=& -2N_g +{16\over 3}s_w^2N_g + 6c_w^2 -n_H({\half}-s_w^2),
\end{eqnarray}
thus leading to the MSSM beta function coefficients
\bege
\Bigg( \matrix{ \b_1 \cr  \b_2 \cr  \b_3 } \Bigg) =
\Bigg( \matrix{ 0 \cr  6 \cr  9 } \Bigg)     -N_g\Bigg( \matrix{ 2 \cr  2 \cr
 2 } \Bigg)  -n_H\Bigg( \matrix{ 3/10 \cr 1/2 \cr  0 } \Bigg).
\ende

The two loop effects contribute to the running of the
couplings through the terms $\theta_i(Q,Q_0)$ in Eq.(\ref{effrun}). Since
we do not have the full mass-dependent contributions, we will have to settle
with the using the usual massless limits. These are explicitly determined by
solving the two-loop renormalization group equations
and are given by
\bege \theta_i(Q,Q_0) =
-{1\over 4\pi}\sum_{j=1}^3{\b_{ij}\over\b_j}
\log{\Big(1+\a_j(Q_0){\b_j\over 4\pi}\log{(Q^2/Q_0^2)} \Big)},
\ende
where the beta matrix is
\bege
\b_{ij}^{MSSM} = -
\Bigg(
 \matrix{ 7.96 & 5.4 & 17.6 \cr 1.8 & 25 & 24 \cr 2.2 & 9 & 14 } \Bigg).
\ende


\newpage

\end{document}